\documentclass[
    aps,
    prx,
    twocolumn,
    10pt,
    a4paper,
    superscriptaddress,
    notitlepage,
    longbibliography,
    floatfix
]{revtex4-2}

\DeclareMathAlphabet      {\mathbfit}{OML}{cmm}{b}{it}

\usepackage{titlesec}

\titlespacing*{\section}{0pt}{2.0ex}{2.0ex}      
\titlespacing*{\subsection}{0pt}{1.5ex}{1.5ex}   

\usepackage{array}
\usepackage{rotating}
\usepackage{placeins}
\usepackage{tabularx}
\usepackage{xcolor}
\definecolor{navyblue}{RGB}{0,0,128}
\definecolor{lightgrey}{gray}{0.65}
\usepackage{amsmath}
\usepackage{multirow}
\usepackage{colortbl,xcolor}
\usepackage{longtable}
\usepackage{graphicx}
\usepackage{bm}
\usepackage{siunitx}
\usepackage{afterpage}
\usepackage{tikz}
\usepackage{hyperref} 
\usepackage{booktabs}
\usepackage{multirow}
\usepackage[normalem]{ulem}

\usepackage{amssymb} 
\newcommand{\yes}{\checkmark}
\usepackage{pifont}
\newcommand{\no}{\ding{55}}

\usepackage{newtxtext,newtxmath}

\usepackage[caption=false]{subfig} 
\usepackage[normalem]{ulem}
\usepackage{xcolor}

\newcommand{\change}[1]{%
  \textcolor{black}{#1}%
}


\usepackage{orcidlink}
\usepackage{hyperref}
\hypersetup{colorlinks=true, citecolor=blue, urlcolor=blue, linkcolor=blue}

\begin{document}

\title{Emergent altermagnetism at surfaces of antiferromagnets:\\
full symmetry classification and material identification
}

\author{Colin Lange \orcidlink{0009-0001-0661-7998}}
\thanks{These authors contributed equally to this work.}
\affiliation{Institute of Physics, Johannes Gutenberg University Mainz, 55099 Mainz, Germany}
\author{Rodrigo Jaeschke-Ubiergo \orcidlink{0000-0002-4821-8303}}
\thanks{These authors contributed equally to this work.}
\affiliation{Institute of Physics, Johannes Gutenberg University Mainz, 55099 Mainz, Germany}
\author{Atasi Chakraborty \orcidlink{0000-0002-5266-6299}}
\affiliation{Institute of Physics, Johannes Gutenberg University Mainz, 55099 Mainz, Germany}
\author{Xanthe H. Verbeek \orcidlink{0000-0003-2262-5047}}
\affiliation{Institute of Physics, Johannes Gutenberg University Mainz, 55099 Mainz, Germany}
\author{Libor Šmejkal \orcidlink{0000-0003-1193-1372}}
 \affiliation{Max Planck Institute for the Physics of Complex Systems, 01187 Dresden, Germany }
    \affiliation{Max Planck Institute for Chemical Physics of Solids, 01187 Dresden, Germany}
\author{Jairo Sinova \orcidlink{0000-0002-9490-2333}}
\affiliation{Institute of Physics, Johannes Gutenberg University Mainz, 55099 Mainz, Germany}
\affiliation{Department of Physics, Texas A\&M University, College Station, Texas 77843-4242, USA}
\author{Alexander Mook \orcidlink{0000-0002-8599-9209}}
\affiliation{University of M\"{u}nster, Institute of Solid State Theory, 48149 Münster, Germany}

\begin{abstract}
We demonstrate the emergence of altermagnetism at the surfaces of antiferromagnets, vastly expanding the number of material candidates with altermagnetic characteristics and establishing a route towards two-dimensional altermagnetism through surface-induced symmetry breaking. We do so by developing a surface spin group formalism that fully classifies all surface magnetic states and identifies altermagnetic surface spin groups that can arise at the surfaces of antiferromagnets. We use this formalism to identify over 150 antiferromagnetic entries from the MAGNDATA database with at least one altermagnetic surface, often times with multiple such surfaces in the same material, and clarify the role of surface roughness and terraces. We illustrate this emergent phenomenon in a realistic Lieb lattice-based minimal model and present ab initio calculations on two representative material candidates, $\text{NaMnP}$ and $\text{FeGe}_2$, exhibiting $d$-wave and $g$-wave surface altermagnetism, respectively. Our theory naturally resolves the contradiction of recent experimental reports of $d$-wave spin splitting from ARPES measurements on metallic Lieb lattice compounds such as KV$_2$Se$_2$O that have been shown to be antiferromagnetic in the bulk. Hence, we establish a new paradigm for generating effectively two-dimensional altermagnetism by functionalizing the abundant material class of collinear antiferromagnets as viable platforms for controlled surface altermagnetism. 
\end{abstract}
\maketitle

\section{Introduction}
\label{Introduction}
The recently discovered altermagnets (AMs) are  magnetically compensated collinear ordered materials, with an unconventional time reversal symmetry (TRS) broken electronic band structure, characterized by an even-wave spin-splitting pattern with $d$-, $g$- or $i$-wave symmetry \cite{Smejkal2021a,Smejkal2022a}. This new unconventional magnetic phase completes the landscape of collinear magnets together with the well-known conventional ferromagnets (FMs) and antiferromagnets (AFMs) \cite{Smejkal2021a,Smejkal2022a,Jungwirth2025,Jungwirth2026}. 
The full classification of bulk collinear magnets in these three types was achieved by considering their spin group symmetries, which allows for distinct symmetries acting on the spin texture and the lattice separately \cite{Litvin1974,Litvin1977}. The symmetry properties of altermagnets lead to several spintronics effects, such as anisotropic spin-polarized currents, anomalous Hall effect, and the novel spin-splitter effect \cite{Jungwirth2025b}. While bulk altermagnetism has been experimentally verified in materials such as $\text{MnTe}$ and $\text{CrSb}$ \cite{Krempasky2024,Lee2024,Osumi2024,Hajlaoui2024,Chilcote2024,Amin2024,Reimers2024,Yang2024,Ding2024,Zeng2024,Li2024,Liu2024b}, \change{and suggested} in the metallic $\text{KV}_2\text{Se}_2\text{O}$ \cite{Jiang2025} and Rb$_{1-\delta}$V$_2$Te$_2$O \cite{Zhang2025a}, there remains an interest in expanding the material class of experimentally readily available altermagnets. This is most relevant in the realm of 2D AMs, since most data technology applications are driven by interface effects. In addition, there is a growing number of interesting effects arising  at altermagnetic interfaces, such as the interplay between altermagnetism and topological superconductivity \cite{Mazin2025,Chatterjee2025}, edge- and cornertronics \cite{Yang2025,Sattigeri2025}, altermagnetic skyrmions \cite{Jin2024, Dou2025}, as well as field switchable ferroelectric altermagnets \cite{Smejkal2024,Gu2025,Wang2024a,Zhu2025,Urru2022} just to mention a few. Most existing strategies to generate 2D AMs focus on exfoliation from bulk AMs  \cite{Mazin2023a,Jana2025,Guo2025,Liu2025a}, bringing an extra step of complication because not all materials are easily exfoliable. 

Given the motivation for 2D altermagnetism, we turn our attention to the largest class of magnetic materials, namely collinear  AFMs -- with about 40\% of materials in the    MAGNDATA material database \cite{Gallego2016a} belonging to this class \cite{Chen2024}. A fundamental question is whether it is possible to functionalize  AFMs in such a way that they generate altermagnetism. Here we demonstrate that this functionalization is possible through the symmetry breaking induced by the termination of particular AFMs in specific directions, i.e., emergent surface-induced altermagnetism in bulk AFMs. 
We do so by developing 
{\it surface spin  groups} and a systematic algorithm to classify all the surfaces of bulk AFMs and identify those having altermagnetic symmetries. This formalism is specific to a semi-infinite geometry with surface states,  living in the 3D real-space and 2D momentum-space, which is different from the spin symmetries in the 3D  \cite{Litvin1977} and strictly 2D monolayer systems \cite{Tian2025a, Zeng2024a}.
By applying this classification to the known collinear AFMs in the MAGNDATA database, we identify over 150 AFMs entries with at least one spin-split altermagnetic surface.  
This new approach gives a  general route to 2D AMs that vastly expands the pool of experimentally accessible AMs, eliminating the need for exfoliation, thereby avoiding the associated material and fabrication challenges.

These altermagnetic states at AFM surfaces can be directly probed by several surface sensitive experimental probes,  such as spin-polarized scanning tunnel microscopy, angle resolved photoemission spectroscopy, and magneto-optical Kerr effect. 
Specifically, the recent reports of neutron scattering experiments  on $\text{KV}_2\text{Se}_2\text{O}$ demonstrating a bulk antiferromagnetic order  \cite{Sun2025a}
and seemingly contradicting the surface-sensitive ARPES altermagnetic measurements \cite{Jiang2025}, are naturally explained by our theory of surface altermagnetism. 
In addition, if these surface altermagnetic states are metallic, they can also exhibit characteristic transport responses, such as the anomalous Hall effect \cite{Smejkal2022AHEReview,Feng2022}, tunneling magnetoresistance \cite{Smejkal2022GMR,Noh2025}, or unconventional spin-polarized and spin-splitter currents.

To illustrate the mechanism of surface-induced altermagnetism, in Sec.~\ref{Toy model},
we first use a minimal tight-binding model of an AFM in a stacking geometry, taking inspiration from materials with layered Lieb-lattice structure. The calculated spin-resolved spectral density reveals directly the altermagnetic character of the surface states. 

In Sec.~\ref{algorithm} we present the
symmetry-based algorithm identifying surface altermagnetism, where we develop the {\it surface spin groups} and their corresponding {\it surface spin Laue groups}, which 
enable the determination and classification of all magnetic surfaces in a given magnetic material. We isolate the surfaces that produce altermagnetic symmetries \change{and explain the effect of surface roughness and terraces}. Applying our methodology to the MAGNDATA database \cite{Gallego2016a}, we identify numerous AFMs as promising candidates that can exhibit surface altermagnetism (SAM). We also indicate AFMs with surfaces that have altermagnetic symmetries but 
where the surface plane coincides with one
of the nodal planes of the altermagnetic order, 
making the surface electronic band structure spin degenerate. We term this scenario spin-degenerate surface altermagnetism (SD-SAM), reserving SAM for the spin-split surface altermagnetism explicitly. \change{To elucidate the physical distinction between SAM and SD-SAM, we analyze in both cases the symmetries giving rise to ferroically-ordered atomic spin densities~\cite{Jaeschke-Ubiergo2025}.} The full list of AFMs extracted from the MAGNDATA database that allow for SAM \change{and/or} SD-SAM is presented in the Appendix, identifying the specific AFMs and the different surfaces with their corresponding magnetic character. 
In Sec.~\ref{Material candidates}, we present explicit ab initio calculation on three representative AFMs identified by our classification, $\text{NaMnP}$\change{, KV$_2$Se$_2$O} and $\text{FeGe}_2$,  which exhibit $d$-wave and $g$-wave spin-split surface altermagnetism (SAM), respectively. 
In Sec.~\ref{discussion} we discuss the implications of our general formalism 
to realize \change{surface} altermagnetism through the controlled termination of bulk AFMs, specifically commenting \change{on current experimental debates on the aforementioned Lieb lattice materials.}
\change{While a quantitative description of structural modifications lies beyond the scope of this work, we briefly comment on how common surface relaxations and reconstructions can influence surface altermagnetism.}
In Sec.~\ref{Conclusion and Outlook} we present our conclusions and outlook on this new approach to engineer surface altermagnetism.

\begin{figure*}
    \centering
    \includegraphics[width=0.8\textwidth]{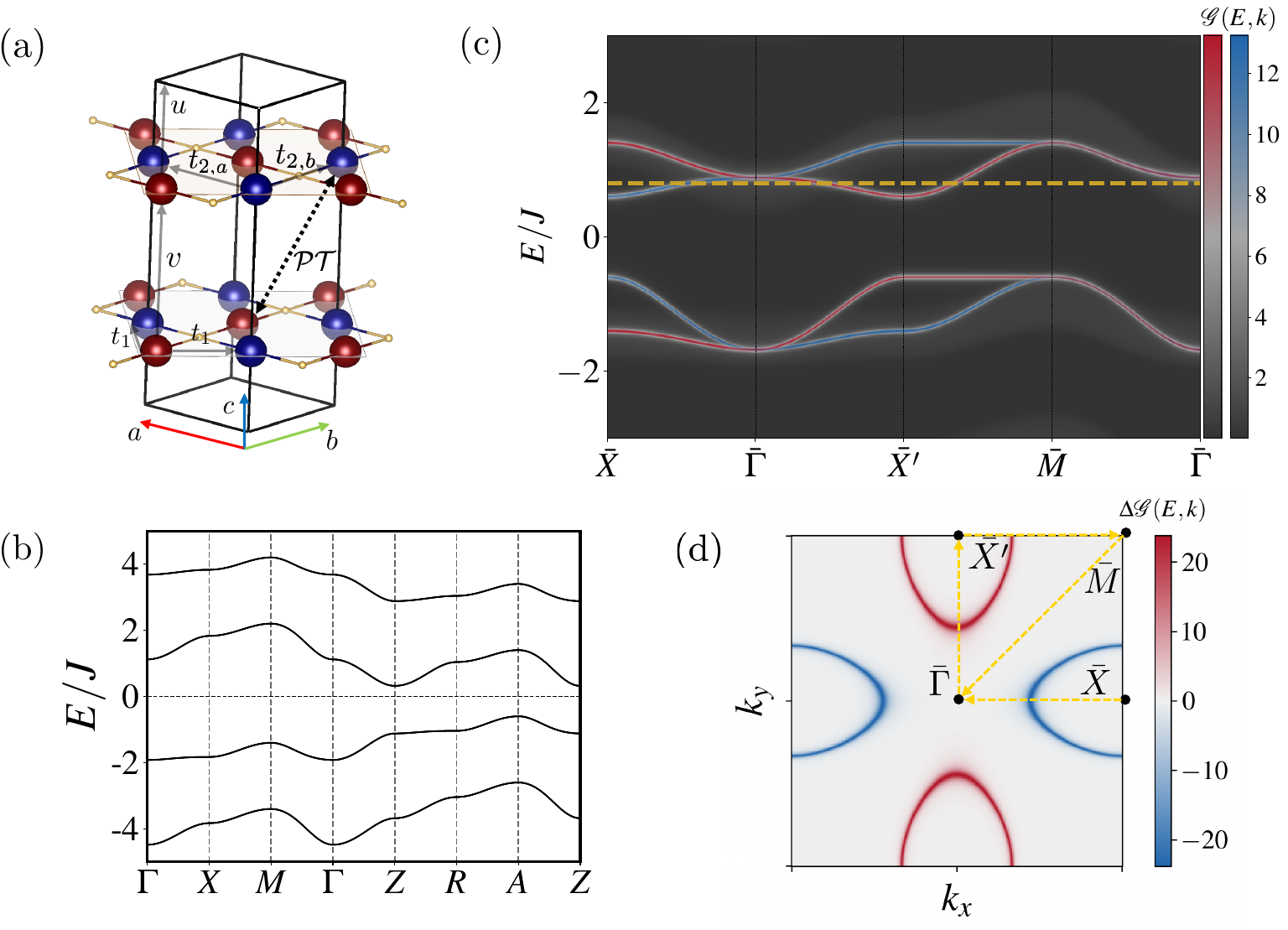}
    \caption{
    Minimal model inspired by the layered Lieb lattice compounds.
    (a) Schematic of the crystal structure for a generic layered Lieb lattice system with the hoppings of the tight-binding model in Eq.~\eqref{toy_hamiltonian} indicated. 
    (b) Kramers' degenarate bulk bands of the minimal model with periodic boundary conditions in all three spatial directions.
    (c) Spin-resolved surface spectral density at the $(001)$ surface showing altermagnetic splitting emerging from the surface-induced symmetry breaking. Red and blue colors depict spectral weight in the spin-up and down channels, respectively. The mixed color signifies the degeneracy of both channels in their spectral density. 
    (d) Isoenergy cut of the difference between the spin-up and down channel of the surface spectral density in the two-dimensional surface Brillouin zone at $0.8/J$ [see dashed yellow line in (c)], exhibiting the $d$-wave pattern of the spin-splitting. Parameters in units of $J$ read: $t_1 = 0.2, t_{2,a}=0.4, t_{2,b} = 0, u = 1.2,J = 1, v = 0.4 $.
    }
    \label{fig:complete_toymodel}
\end{figure*}

\section{Minimal model for emergent surface altermagnetism}
\label{Toy model}

The inverse Lieb lattice \cite{Kaushal2024, Duerrnagel2024,Chang2025} has attracted considerable attention as a minimal and conceptually clean platform to achieve two-dimensional $d$-wave altermagnetism. In the first report of two-dimensional altermagnetism, a single monolayer of FeSe \cite{Mazin2023a}, a parent compound of extensively studied FeSe-based superconductors, was shown to recover the same checkerboard order with altermagnetic symmetry, after breaking the inversion center that connects the opposite-spin Fe sites. In the same work, the magneto-optical Kerr effect was calculated as a probe of the time-reversal symmetry breaking at the surface of an FeSe slab system. Additionally, recent experimental reports on the metallic Lieb lattice compounds 
Rb$_{1-\delta}$V$_2$Te$_2$O \cite{Zhang2025a}
and KV$_2$Se$_2$O \cite{Jiang2025} suggest altermagnetic order with $d$-wave symmetry at room temperature, 
but may in fact be explained by surface altermagnetism, as suggested by recent neutron scattering measurements indicating bulk antiferromagnetic order \cite{Sun2025a}.
Both materials are of particular relevance to our discussion because the altermagnetic spin splitting was probed by spin- and angle-resolved photoemission spectroscopy, which directly accesses the states at the surface of the sample.

In the aforementioned family of materials, the bulk magnetic order depends on the stacking sequence of consecutive Lieb-lattice layers. When identical magnetic layers are stacked directly on top of each other, as was assumed in the reported experiments of KV$_2$Se$_2$O \cite{Jiang2025} and Rb$_{1-\delta}$V$_2$Te$_2$O \cite{Zhang2025a}, the system is altermagnetic. 
By contrast, when the layers are stacked with alternating crystal environments, as in the representative structure shown in Fig.~\ref{fig:complete_toymodel}(a), the resulting system recovers the symmetries of an antiferromagnet.

We borrow intuition from this material class, and we write a minimal model that illustrates the symmetry-breaking mechanism behind the surface-induced altermagnetism. Later, in Sec.~\ref{sec:NaMnP}, we will demonstrate the same effect in a closely related structure, in the antiferromagnetic NaMnP \cite{Bronger1986}. 
Our tight-binding model depicted in Fig.~\ref{fig:complete_toymodel}(a) consists of coupled layers, with each layer being a prototypical two-dimensional $d$-wave AM with checkerboard pattern. Red and blue spheres indicate the magnetic sites with opposite magnetic moment, characterized in the model by a Hund's coupling constant $J$. The altermagnetism in a single layer emerges from the presence of the nonmagnetic sites (denoted by a yellow sphere) which break the inversion center between opposite-spin sublattices within the same layer. Focusing on a given sublattice (red or blue), the intra-sublattice bonds along the $[100]$ and $[010]$ directions are inequivalent due to the presence of a nonmagnetic ligand along only one of these directions. 

We integrate out the degrees of freedom associated with the yellow ligands, which in real materials can occupy a different Wyckoff position, and we introduce the resulting sublattice anisotropy through intra-sublattice hopping amplitudes $t_{2,a}$ and $t_{2,b}$, with $t_{2,a}\neq t_{2,b}$. These intra-sublattice hoppings alternate between adjacent layers, leading to a conventional antiferromagnetic order. Specifically, this alternation introduces the spin symmetry  $[\mathcal{T}||\mathcal{PT}]$, which we abbreviate as $\mathcal{PT}$, with $\mathcal{P}$ and $\mathcal{T}$ denoting the real-space inversion, and the time-reversal symmetry, respectively. This $\mathcal{PT}$ symmetry protects the spin degeneracy across the entire Brillouin zone. Here, the double bar $||$ separates the spin and real space operations. We also introduce a nearest neighbor inter-sublattice hopping amplitude $t_1$, which is identical in every layer.

We wish to engineer surface states that preserve the altermagnetic properties of a single layer at the antiferromagnet’s surface. To this end we introduce alternating vertical inter‑layer couplings $u$ and $v$, i.e., an SSH‑type staggering of the hopping amplitudes \cite{Su1979}. The SSH inter-layer couplings $u\neq v$ break the center of inversion at the nonmagnetic ligand. Because of this, the real space inversion only connects opposite-spin sublattices [see Fig.\ref{fig:complete_toymodel}(a)], and $[E||\mathcal{P}]$ is not a symmetry. However, the spin-only group of a collinear magnet, $\mathbf{r}_{\rm so}= \mathrm{SO}(2)\rtimes\mathbb{Z}_2^{[C_{2\perp}\mathcal{T}||\mathcal{T}]}$, introduces an effective inversion in the electronic band structure \cite{Smejkal2021a}. Here, $C_{2\perp}$ is a two-fold spin rotation around an axis perpendicular to the collinear axis of the spins and $E$ denotes the identity.
The corresponding spin point group after introducing inversion is called the spin \change{Laue} group, and for this model reads 
\begin{equation}
    \mathbb{Z}_2^{[C_{2\perp}||E]}\times
    {}^14/^1m^1m^1m,    
\end{equation}
where $\mathbb{Z}_2^{[C_{2\perp}||E]}=\{[E||E], [C_{2\perp}||E]\}$. We adopt the notation introduced by Litvin \cite{Litvin1974, Litvin1977}. In this convention, the symbol $^14/^1m^1m^1m$ denotes the spin point group constructed from the crystallographic point group $4/mmm$ (Hermann–Mauguin notation) by associating a spin operation with each real-space symmetry generator. The superscript preceding each symbol specifies the corresponding spin operation. In particular, superscripts 1 and 2 denote the identity operation and a two-fold spin rotation about an axis perpendicular to the collinear spins, respectively.

The resulting Bloch Hamilton matrix is given by
\begin{align}\label{toy_hamiltonian}
    \begin{split}
        \mathcal{H}_{\boldsymbol{k}} = &\change{-(v+u)\cos \left(k_z/2\right)\sigma_x \otimes\tau_0\otimes\nu_0}\\
        &\change{-(v-u)\sin\left(k_z/2\right)\sigma_y\otimes\tau_0\otimes\nu_0}\\
        &- t'(\boldsymbol{k})\,\sigma_0\otimes\tau_x\otimes\nu_0- J\,\sigma_0\otimes\tau_z\otimes\nu_z\\
        &- \big(\mathcal{A}(\boldsymbol{k})-\mathcal{B}(\boldsymbol{k})\big)\,\sigma_z\otimes\tau_z\otimes\nu_0\\
        &- \big(\mathcal{A}(\boldsymbol{k})+\mathcal{B}(\boldsymbol{k})\big)\,\sigma_0\otimes\tau_0\otimes\nu_0,
    \end{split}
\end{align}
where $\sigma_\mu$, $\tau_\mu$, and $\nu_\mu$ with $\mu \in \{ 0,x,y,z \}$ are the Pauli matrices denoting the layer, the sublattice, and the spin degree of freedom of the electrons, respectively. We have further defined
\begin{align}
    \begin{split}
        \mathcal{A}(\boldsymbol{k}) &= t_{2,b}\cos\left(k_x\right)+t_{2,a}\cos\left(k_y\right), \\
        \mathcal{B}(\boldsymbol{k}) &= t_{2,a}\cos\left(k_x\right)+t_{2,b}\cos\left(k_y\right), \\
        t'(\boldsymbol{k}) &= t_{1}\cos\left(\frac{1}{2}(k_x+k_y)\right)+t_{1}\cos\left(\frac{1}{2}(k_y-k_x)\right).
    \end{split}
\end{align}
The corresponding electronic band structure of this bulk AFM model is shown in Fig.~\ref{fig:complete_toymodel}(b). The presence of the $\mathcal{PT}$ symmetry enforces the Kramers' degeneracy of the bands.

To resolve the states at the $(001)$ \textit{surface} of a semi-infinite version of this model, we use the established surface Green's function renormalization method \cite{LopezSancho1984, LopezSancho1985, Henk1993}. It provides access to the spin-resolved spectral function $\mathcal{G}_\sigma(E,\boldsymbol{k}) = - \text{Im} \text{Tr} G_\sigma(E,\boldsymbol{k}) / \pi$ where $G_\sigma(E,\boldsymbol{k})$ is the surface Green's function with spin $\sigma =~\uparrow,~\downarrow$. In Fig.~\ref{fig:complete_toymodel}(c), we plot the overlaid  $\mathcal{G}_\uparrow(E,\boldsymbol{k})$ and $\mathcal{G}_\downarrow(E,\boldsymbol{k})$ along a high-symmetry momentum path in the surface Brillouin zone. One can clearly see the high spectral weight of surface states with alternating spin polarization. The $d$-wave spin splitting in the surface spectrum becomes particularly clear for the isoenergy cut of $\Delta \mathcal{G}(E,\boldsymbol{k}) = \mathcal{G}_\uparrow(E,\boldsymbol{k}) - \mathcal{G}_\downarrow(E,\boldsymbol{k})$ shown in Fig.~\ref{fig:complete_toymodel}(d)

The spin Laue group of the bulk,
\[   \mathbb{Z}^{[C_{2\perp}||E]}_2\times {}^14/^1m^1m^1m=[E||4/mmm]+[C_{2\perp}||4/mmm],
\] 
is broken down to 
\[
^24/^1m^1m^2m=[E||mmm]+[C_{2\perp}||4/mmm-mmm]
\]
by the termination of the $(001)$ surface. Here, $4/mmm-mmm$ denotes the set difference between groups $4/mmm$ and $mmm$. Note that both groups are centrosymmetric (i.e., they are spin Laue groups), despite the fact that inversion symmetry is broken by the surface. This is because the collinearity of the spins induces an effective inversion in the electronic band structure, originating from the spin-only group \cite{Smejkal2021a}.

The spin point group
$^24/^1m^1m^2m$ corresponds to a $d$-wave altermagnet \cite{Smejkal2021a} (and reciprocal space group 6 from Ref.~\cite{Zeng2024a}). The symmetry $[C_{2\perp}||C_{4z}]$, with $C_{4z}$ a four-fold rotational axis parallel to $z$ that passes through the nonmagnetic ligand [yellow sphere in Fig.~\ref{fig:complete_toymodel}(a)], connects opposite-spin sublattices, and is reflected in the sign change of the spin-polarized surface spectral function under a 90° rotation [see path $\bar{X}\bar{\Gamma} \bar{X}'$ in Fig.~\ref{fig:complete_toymodel}(c)]. Additional sublattice transposing symmetries are the mirrors $[C_{2\perp}||m_{[110]}]$ and $ [C_{2\perp}||m_{[1\bar{1}0]}$]. These transposing mirrors enforce two spin-degenerate nodal planes, which become nodal lines in the projected 2D surface Brillouin zone [see $\bar{\Gamma} \bar{M}$ path in Fig.~\ref{fig:complete_toymodel}(c,d)].

This minimal model, inspired by the broad family of materials with a layered structure of Lieb lattices, aims to illustrate a simple mechanism of how altermagnetism can emerge at the surface of an antiferromagnet: The symmetries enforcing spin degeneracy throughout the bulk Brillouin zone (here: $\mathcal{PT}$ symmetry) are broken by the surface termination of the AFM, while some spin symmetries that protect the magnetic compensation are preserved.

\section{Systematic search for surface altermagnetism}
\label{algorithm}

A natural next step is to develop a systematic strategy for identifying real materials that host surface altermagnetism. Establishing the appropriate mathematical framework, together with an algorithmic approach for its detection, serves two complementary purposes. First, it enables a complete classification of all symmetry scenarios in which surface altermagnetism can, in principle, arise. Second, it allows us to perform database searches of magnetic materials to identify concrete candidates and to compile statistics on the prevalence of bulk antiferromagnets exhibiting surface altermagnetism.
We will proceed in three steps: (i) we define and discuss the mathematical objects of \textit{surface} spin groups in Sec.~\ref{sec:surfacegroups}. (ii) We classify surface spin groups describing altermagnetism in Sec.~\ref{sec:classification}. (iii) We develop an algorithm to identify bulk AFM with surface altermagnetism in the MAGNDATA materials database in Sec.~\ref{sec:algorithm}. \change{(iv) We explain how the influence of surface roughness and terraces on the altermagnetic surface splitting can be inferred from the bulk spin space group in Sec.~\ref{sec:terraces}.} Finally, (v) we present its results to demonstrate the abundance of surface altermagnets in Sec.~\ref{sec:screening-results}.

\subsection{Definition of surface spin groups} \label{sec:surfacegroups}

We introduce the notion of a \textit{surface} spin space group (sSSG), a notion that extends the spin-group framework \cite{Litvin1974,Litvin1977} to semi-infinite systems -- similar to what the magnetic surface group formalism in Ref.~\cite{Weber2024} provided for magnetic groups. We propose that the sSSG is the natural object for investigating exchange-dominated physics in experiments that resolve electronic states localized at the surface of the sample.

We define the nontrivial sSSG with respect to a nontrivial bulk SSG $R^s$ and a surface normal $\boldsymbol{n}$ as 
\begin{align}\label{sSSG}
    R^s_{\boldsymbol{n}} = \{[g_s||g_r|\boldsymbol{t}]\in R^s: (g_r \cdot \boldsymbol{n} = \boldsymbol{n} ) \land ( \boldsymbol t \perp \boldsymbol{n} ) \},
\end{align}
where $g_s$ and $g_r$ are point symmetries acting on spin and real space, respectively, and $\boldsymbol{t}$ is a translation in direct space. Therefore, $[g_s||g_r|\boldsymbol{t}]$ is an element of the nontrivial bulk SSG $R^s$. For brevity, we will refer to $R^s_{\boldsymbol{n}}$ as the sSSG and drop the prefix ``nontrivial''; this is justified by the fact that the spin-only group is always unaffected by the surface in real space. 
Note that $R^s_{\boldsymbol{n}}$ is a subgroup of $R^s$. Specifically, $R^s_{\boldsymbol{n}}$ comprises all elements of the bulk SSG that keep the normal $\boldsymbol{n}$ of a given surface invariant and either (i) have no translational component at all, or (ii) involve translations purely within the surface plane (normal to $\boldsymbol{n}$). 

In the remainder of this work, we restrict our analysis to spin \textit{point} group elements within the surface spin point group (sSPG). To obtain those from the elements of the sSSG, we project out the translational part of each symmetry operation, identifying it with the corresponding point group transformation. We use the symbols $\mathcal{R}^s$ and $\mathcal{R}^s_{\boldsymbol{n}}$ 
to denote the spin point group and sSPG associated with the bulk spin space group $R^s$ and sSSG $R^s_{\boldsymbol{n}}$, respectively.

In order to appreciate in what way the sSPG differs from the more familiar SPG in 3D systems \cite{Litvin1977} and strict 2D monolayer systems \cite{Tian2025a, Zeng2024a}, we recapitulate how the latter two are constructed. There, one employs three- or two-dimensional representations in both real \textit{and} reciprocal space. In contrast, since sSPG are meant to describe a semi-infinite system, the real space operations remain three-dimensional, but the Brillouin zone is projected onto only two dimensions, i.e., the surface Brillouin zone. 

Next, we explain how this difference enters on a technical level. In 3D systems, the spin point group element $[C_{2\perp}||E]$ enforces spin degeneracy across the entire Brillouin zone \cite{Smejkal2021a}. Additionally, for collinear systems, the symmetry $[C_{2\perp}||\mathcal{P}]$ together with $[C_{2\perp}\mathcal{T}||\mathcal{T}]\in \mathbf{r}_{\rm so}$ \cite{Litvin1974} implies the presence of $[\mathcal{T}||\mathcal{PT}]$ that enforces spin degeneracy even when SOC is included, because of Kramers' theorem \cite{Kramers1930, Wigner1932, Smejkal2021a}.
In the strict 2D limit, there are two additional elements, $[C_{2\perp}||m_{\boldsymbol{n}}]$ and $[C_{2\perp}||C_{\boldsymbol{n}}]$, which analogously enforce spin-degeneracy for collinear magnets \cite{Tian2025a, Zeng2024a}. 
However, for semi-infinite systems with a surface, only $[C_{2\perp}||C_{\boldsymbol{n}}]$ can be part of the nontrivial sSPG because $[C_{2\perp}||m_{\boldsymbol{n}}]$ and $[C_{2\perp}||\mathcal{P}]$ by definition do not leave the normal of the respective surface invariant.
This is the central distinguishing feature of the sSPG from both the well-established 3D SPG \cite{Smejkal2021a} and the 2D spin layer point groups \cite{Zeng2024a}: Only the elements $[C_{2\perp}||E]$ and $[C_{2\perp}||C_{2\boldsymbol{n}}]$ give rise to spin-degenerate bands (the latter assuming collinear spins), placing sSPG in an intermediate regime between the fully three- and two-dimensional limits.

\begin{table*}
\caption{Classification of altermagnetic surface spin point groups. The first column specifies the partial-wave character associated with the surface spin group, that can be $d$-, $g$-, $i$-wave \cite{Smejkal2021a} or spin degenerate (SD-SAM).
The second and third columns list the surface spin point group  $\mathcal{R}_{\boldsymbol{n}}^s$ and the corresponding surface spin Laue group $\mathbb{Z}_2^{[E||\mathcal{P}]}\!\times\!\mathcal{R}_{\boldsymbol{n}}^s$, respectively. These spin groups are given in Litvin notation~\cite{Litvin1977}. The coordinate system is chosen such that $z$ is parallel to the surface normal and $x,y$ span the surface plane. The fourth column indicates whether the corresponding group generates spin splitting in the 2D surface Brillouin zone. A tick (\yes) indicates spin-split surface bands. These groups describe surface altermagnetism (SAM). In contrast, a cross (\no) denotes the absence of spin-splitting in the non-relativistic limit. These groups describe spin-degenerate surface altermagnetism (SD-SAM).
The final column reports the number of surfaces of the respective symmetry group identified in our materials search of the MAGNDATA database (see Tab.~\ref{tab:table}). For consistency with the color-code used in Tab.~\ref{tab:table}, those surface spin Laue groups in the third column that lead to spin-split surface states (SAM) are written in blue.
}
\label{tab:tabsummary}

\begin{ruledtabular}
\renewcommand{\arraystretch}{1.35}
\begin{tabular}{ c  c  c  c  c }
Surf. partial-wave character &
Surf. spin point group 
&
Surf. spin Laue group 
&
Spin split surf. states? &
No.\ of surfaces \\

\hline
$d$-wave
& $^{2}m_x\,^{2}m_y\,^{1}2_z$
& \textcolor{blue}{$^{2}m_x\,^{2}m_y\,^{1}m_z$}
& \yes & \change{145} \\
& $^{2}4\,^{2}m\,^{1}m$
& \textcolor{blue}{$^{2}4/^{1}m\,^{2}m\,^{1}m$}
& \yes & \change{82} \\
& $^{2}4$
& \textcolor{blue}{$^{2}4/^{1}m$}
& \yes & 1 \\

& $^{2}m_{x/y}$
& \textcolor{blue}{$^{2}2_{x/y}/^{2}m_{x/y}$}
& \yes & \change{64} \\

\hline
$g$-wave
& $^{1}4\,^{2}m\,^{2}m$
& \textcolor{blue}{$^{1}4/^{1}m\,^{2}m\,^{2}m$}
& \yes & \change{4} \\

\hline
$i$-wave
& $^{1}6\,^{2}m\,^{2}m$
& \textcolor{blue}{$^{1}6/^{1}m\,^{2}m\,^{2}m$}
& \yes & -- \\
& $^{1}3\,^{2}m$
& \textcolor{blue}{$^{1}\bar{3}\,^{2}m$}
& \yes & \change{2} \\

\hline
SD-SAM
& $^{2}m_{x/y}\,^{1}m_{y/x}\,^{2}2_z$
& $^{2}m_{x/y}\,^{1}m_{y/x}\;^{2}m_z$
& \no & \change{306} \\
& $^{2}2_z$
& $^{2}2_z/^{2}m_z$
& \no & \change{66} \\
& $^{2}6\,^{2}m\,^{1}m$
& $^{2}6/^{2}m\,^{2}m\,^{1}m$
& \no & -- \\
& $^{2}6$
& $^{2}6/^{2}m$
& \no & 1 \\
\end{tabular}
\end{ruledtabular}
\end{table*}

\subsection{Classification of altermagnetic surface spin groups} \label{sec:classification}

As in 3D and 2D spin groups, only specific surface spin groups support altermagnetism. These groups can be further classified according to the nodal structure they impose on the spin splitting of the altermagnetic surface, yielding $d$-, $g$-, and $i$-wave types characterized by two, four, and six nodal lines, respectively. In contrast to three-dimensional spin groups -- but in close analogy with two-dimensional spin groups~\cite{Bai2025a} -- surface spin groups occur in two distinct variants, depending on whether the surface coincides with a nodal plane of the altermagnetic order. When the surface does not coincide with a nodal plane, the surface electronic spectrum is spin split, as expected for altermagnets; we refer to this case as surface altermagnetism (SAM). If, however, the surface lies within a nodal plane, the surface spin group still exhibits altermagnetic symmetries, yet the surface spectrum remains spin degenerate. We denote this latter scenario as spin-degenerate surface altermagnetism (SD-SAM). 

\change{The SD-SAM scenario is reminiscent of the type-IV magnet scenario discussed in Ref.~\cite{Bai2025a} because both are properties present in a 2D BZ. In SD-SAM, however, the relevant two-dimensional Brillouin zone is not that of an isolated layer, but the \textit{surface} Brillouin zone of a semi-infinite three-dimensional crystal. Thus, although the symmetry constraints appear closely related, they apply to distinct physical settings.}

\change{As we will show, even though surfaces with SD-SAM exhibit no spin polarization in momentum space in the non-relativistic limit, they still have the defining altermagnetic symmetries: collinearity, symmetry compensation of magnetic moments, no $\mathcal{PT, T}\boldsymbol{t} $ symmetries. These properties are reflected in the presence of ferroically ordered partial-wave components of the atomic spin density (see Ref.~\cite{Jaeschke-Ubiergo2025}). This ferroic order, which is constrained by the altermagnetic spin symmetries, breaks time reversal symmetry while retaining a net zero magnetization. However, for SD-SAM, the ferroic atomic spin densities have one of their nodal planes coinciding with the surface, leading to zero spin splitting in the nonrelativistic limit. In Appendix \ref{spin_density_SDAM_vs_SAM}, we provide a more detailed comparison of the ferroic order in SAM and SD-SAM.}



\change{The symmetries responsible for the non-relativistic spin degeneracy in the band structure of SD-SAMs are absent in the corresponding magnetic point group, thereby revealing the symmetry-enforced higher-parity-wave spin-splitting pattern characteristic of altermagnets. Consequently, many defining transport properties associated with altermagnetism, such as the magneto-optical Kerr effect (MOKE) or the anomalous Hall effect (AHE), remain present in SD-SAMs. This follows from the fact that these transport phenomena are constrained by the magnetic point group rather than by the spin group, within which SD-SAMs and SAMs are indistinguishable.}

The full classification of surface altermagnetism is presented in Tab.~\ref{tab:tabsummary}. In the following, we will explain this table as we fill in the technical details.

Recall the definition of $\mathcal{R}_{\boldsymbol{n}}^s$ as the nontrivial sSPG 
associated with the nontrivial surface spin space group $R_{\boldsymbol{n}}^s$ in Eq.~\eqref{sSSG}. Again, for brevity, we will drop the prefix ``nontrivial'' from here on when it is clear which group we mean. 
In order to identify altermagnetic sSPG,
we first investigate the spatial parent point groups without spin information, i.e., the projection of the sSPG on the spatial part. In the following, when we say ``groups'', we imply that we are talking about point groups.

Our strategy to end up with a full classification of all spin \textit{Laue} groups starts by eliminating from the 32 crystallographic point groups all groups that are \textit{not} viable surface point groups. After that we analyze which spin Laue groups can be produced by the remaining spatial point groups. We do so by adding the effective inversion symmetry of the collinear spin-only group back to the nontrivial sSPG to generate the corresponding surface spin Laue group. As a final step, we discuss which of the surface spin Laue groups belong to the SAM class or to the SD-SAM class, based on their action on the electronic band structure.

We realize that all centrosymmetric groups and those generated by a roto-inversion (such as $\bar{4}$) cannot be a spatial parent of a sSPG because all inversions (and rotoinversions) do not leave any normal invariant. The second restriction is that there ought to be only one rotation axis present in the surface point group, because any two perpendicular rotation axes would not leave any normal invariant. For example, this restriction renders the point group $222$ an impossible parent group for surface groups. Likewise, the identity trivially cannot support SAM or SD-SAM and is thus also not included. Taking these points into consideration, we are left with the following 9 candidates for spatial parent groups for the sSPG: 
\begin{align}
 \{2, m, mm2, 4, 4mm, 3, 3m, 6, 6mm\}.   \label{eq:candidate-parents}
\end{align}

It is now necessary to add the spin information back because the directions of the symmetry axis associated with compensating symmetries are essential to distinguish between the SAM and SD-SAM classes. We note that in principle all groups in Eq.~\eqref{eq:candidate-parents} can also support an \textit{antiferromagnetic} surface if the symmetry $[C_{2\perp}||E]$ is present at the surface. As before, this arises from the projection of a spin translation element $[C_{2\perp}||E|\tau]$. At the level of the surface spin Laue group, this then produces the generic form $\mathbb{Z}_2^{[C_{2\perp}||E]}\times \mathcal{G}$, where $\mathcal{G}$ is one of the point groups mentioned above. Because in this manuscript we focus on altermagnetic surfaces, we will not discuss this situation further.

Without loss of generality, and for notational simplicity, we assume throughout this discussion that the surface normal is along the $z$ direction, with the $x$ and $y$ directions lying in the surface plane. Following the convention in Ref. \cite{Smejkal2021a}, we choose $\{E, C_{2\perp}\}$ as the spin parent group for the construction of the collinear spin groups. Because this spin parent group has two elements, every subgroup of the full product group between the spin and the spatial parent group can be built from halving subgroups of the spatial parent. This is the particular case of Litvin's theorem for binary spin parent groups \cite{Litvin1974, Litvin1977}.

We now move to different combinations of the aforementioned spin parent group with the 9 spatial parent point groups in Eq.~\eqref{eq:candidate-parents} and go through the list one-by-one: 

(1) We start with the group $2_z$. (We indicate the $z$ in the subscript because we have chosen the $z$ axis as the surface normal.) This group only has the trivial subgroup that generates the nontrivial sSPG $^22_z$. 

(2) The same argument applies to the group $m$ where now the normal has to be inside the mirror plane, generating $^2m_{x/y}$. The parent point group $m_z$, however, is \textit{not} eligible because it does not leave the normal in $z$ invariant. From here on, we will \textit{not} mention orientations of symmetry elements that are forbidden due to the restriction of the sSPG and instead only discuss cases that actually work.

(3) The group $mm2$ has three viable halving subgroups, namely, $m_x$, $m_y$, and $2_z$. So, we can have two distinct nontrivial sSPG from this; they are $^2m_{x/y}\,^2m_{y/x}\,^12_z$ and $^2m_{x/y}\,^1m_{y/x}\,^22_z$, where we have again used the fact that the rotation axis must be parallel to the normal. 

(4) The group $4$ has only one halving subgroup, $2$, so we get only one nontrivial sSPG: $^24_z$. 

(5) The group $4mmm$ has $4$ and $mm2$ as halving subgroups, generating the two nontrivial spin point groups $^24^1m^2m$ ($\cong \, ^24^2m^1m$) and $^14^2m^2m$, where again, only one rotation axis is allowed, which must be parallel to the normal of the surface. 

(6) The group $3$ is a special case because it does not have a halving subgroup. Thus it cannot support AM or SD-SAM and is thus ruled out. 

(7) In contrast, $3m$ has only one halving  subgroup, which is $3$ and generates the nontrivial sSPG $^13^2m$. 

(8) The group $6$ only has $3$ as a halving subgroup, generating the nontrivial sSPG $^26$.

(9) Finally, the group $6mm$, similar to $4mm$, has one halving subgroup generated by the rotations and two isomorphic ones generated by the mirror. Thus, we get the nontrivial sSPG $^26^2m^1m$ and $^16^2m^2m$. 

We conclude that there are in total the following 11 different nontrivial sSPG:
\[
\left\{
\begin{aligned}
&{}^{2}2_z,\; {}^{2}m_{x/y},\; {}^{2}m_{x/y}{}^{2}m_{y/x}{}^{1}2_z,\; {}^{2}m_{x/y}{}^{1}m_{y/x}{}^{2}2_z,\; {}^{2}4_z, \\
&{}^{2}4^{1}m^{2}m,\; {}^{1}4^{2}m^{2}m,\; {}^{1}3^{2}m,\; {}^{2}6,\; {}^{2}6^{2}m^{1}m,\; {}^{1}6^{2}m^{2}m
\end{aligned}
\right\},
\]
which are listed in the second column of Tab.~\ref{tab:tabsummary}.

We now add the effective inversion symmetry of the spin-only group to build surface spin Laue groups from this. This step produces the following list:
\[
\left\{
\begin{aligned}
&{}^{2}2_z/{}^{2}m,\; {}^{2}2/{}^{2}m_{x/y},\; {}^{2}m_{x/y}{}^{2}m_{y/x}{}^{1}m_z, \\
&{}^{2}m_{x/y}{}^{1}m_{y/x}{}^{2}m_z,\; {}^{2}4_z/{}^{1}m,\; {}^{2}4_z/{}^{1}m^{1}m^{2}m, \\
&{}^{1}4_z/{}^{1}m^{2}m^{2}m,\; {}^{1}\bar{3}^{2}m,\; {}^{2}6/{}^{2}m,\; {}^{2}6/{}^{2}m^{2}m^{1}m,\; {}^{1}6/{}^{1}m^{2}m^{2}m
\end{aligned}
\right\}.
\]
These 11 surface spin Laue groups of the form $\mathbb{Z}^{[E||\mathcal{P}]}\times\mathcal{R}_{\boldsymbol{n}}^s$ correspond to the 11 groups listed in the third column of Tab.~\ref{tab:tabsummary}.

In the first column of Tab.~\ref{tab:tabsummary}, we list the partial-wave character, constrained by the surface spin Laue group. We order the possible groups as $d$-, $g$-, and $i$-wave \cite{Smejkal2021a}. When analyzing the spin-splitting symmetry that each of these groups imposes, one should note that, due to the 2D nature of the projected Brillouin zone (surface Brillouin zone), in the spin split cases (see 4th column in Tab.~\ref{tab:tabsummary}), there will be spin-degenerate nodal \textit{lines} instead of nodal planes. All of them are forced to be straight lines, with the only exception of the spin Laue group $^24/^1m$, in which the symmetry $[C_{2\perp}||C_{4z}]$ enforces two spin unpolarized curves (``nodal curves'') that cross the $\Gamma$ point, but do not have to be straight. The surface partial-wave character of almost all surface spin Laue groups, is directly inherited from the 3D spin point groups \cite{Smejkal2021a}, with two exceptions: (i) the SD-SAM cases, discussed in the next paragraph; and (ii) the group ${}^1 \bar{3} {}^2m$, which in three dimensions exhibits $g$-wave symmetry, characterized by three spin degenerate nodal planes and one generally curved nodal surface. Upon projection to two dimensions, this partial-wave structure yields an $i$-wave spin-splitting pattern with six nodal lines: three originating from the nodal planes and three arising from the intersection of the curved nodal surface with the $k_z=0$ plane \cite{Mazin2023a}.

Regarding the question of the groups describing SAM or SD-SAM, we focus on the compensating two-fold rotation axis $[C_{2\perp}||C_{2\boldsymbol{n}}]$ because it generates a compensating \textit{mirror} in the plane of the surface when considering the associated Laue group. Since the Brillouin zone gets projected onto two dimensions (i.e., onto the surface Brillouin zone spanned in the $xy$-plane) we get under the action of the spin Laue group element
\[
[C_{2\perp}||m_z] \epsilon(s,k_x,k_y) = \epsilon(-s,k_x,k_y) \stackrel{!}{=} \epsilon(s,k_x,k_y),
\]
which implies spin degeneracy of the bands $\epsilon(s,k_x,k_y)$, where $\epsilon$ denotes energy and $s$ spin. The groups that have (that do \textit{not} have) this spin-group element are labeled with ``\no'' (``\yes'') in the fourth column of Tab.~\ref{tab:tabsummary}. The checkmark and the cross therefore, indicate if the electronic surface spectrum is spin split (\yes) or not (\no).

Note that although SD-SAM surfaces do not display the characteristic alternating spin splitting in momentum space in the non-relativistic limit, they do \textit{not} have time reversal as a point symmetry (as opposed to \textit{antiferromagnetic} surfaces). Despite the 2D nature of the surface Brillouin zone, in real space, the spin density is still a three-dimensional object, which implies that there is a ferroic order of the partial-wave components of the atomic spin density, with characteristic $d$- or $g$-wave symmetry \cite{Fernandes2023, McClarty2024, Jaeschke-Ubiergo2025}. SD-SAM cannot have $i$-wave symmetry.

\subsection{Search algorithm for surface altermagnets} \label{sec:algorithm}

With all altermagnetic surface groups classified (see Sec.~\ref{sec:classification} and Tab.~\ref{tab:tabsummary}), we now turn to the question of how to identify the surface spin Laue group of a specific surface of a specific material. To develop such an algorithm, first, note that choosing a generic crystal direction as a normal $\boldsymbol{n}$ of the surface will only generate the trivial sSSG, which is not altermagnetic, because it lacks any compensating symmetry; instead, it should be considered a ferromagnetic, uncompensated surface. We therefore have to apply filters as to what normals are sensible candidates.

We restrict ourselves to surfaces whose normals are symmetry axes of the bulk SSG, that is, normals that are left invariant by certain operations of the bulk SSG. 
For example, consider again the minimal model in Sec.~\ref{Toy model} and investigate it from the perspective of the sSSG. As stated in Sec.~\ref{Toy model}, the symmetry group of the bulk system is given by $\mathbb{Z}^{[C_{2\perp}||E]}_2\times\vspace{1pt}^14/^1m^1m^1m$. Consider terminating the bulk of the minimal model in the $\boldsymbol n \parallel [001]$ direction. Clearly, the sublattice-transposing fourfold rotation and the orthogonal mirrors (which come both as sublattice-transposing and nontransposing) leave $\boldsymbol n$ invariant. Thus, the corresponding nontrivial sSPG of the bulk system terminated in the $[001]$ direction is $^24^1m^2m$. If we then go to the surface spin Laue group, by adding the effective inversion of the spin-only group, we arrive at $^24/^1m^1m^2m$. According to the third row in Tab.~\ref{tab:tabsummary}, we conclude that the $[001]$ surface of the minimal model is a $d$-wave SAM surface. The expected $d$-wave spin splitting is evident from the explicit calculations carried out in Sec.~\ref{Toy model} (recall Fig.~\ref{fig:complete_toymodel}). 
Other surface normals of the minimal model that are left invariant under some bulk SSG operations are $[100]$ and $[010]$. Their sSPG is $^1m^2m^22_{\boldsymbol{n}}$. This group, and thus the corresponding surface spin Laue group $^1m^2m^2m$, has the element $[C_{2\perp}||C_{2\boldsymbol{n}}]$. Therefore, these are SD-SAM surfaces (see second row in Tab.~\ref{tab:tabsummary}).

We now take this intuition from the minimal model and turn to the systematic analysis of \textit{all} conventional collinear 3D antiferromagnets in the MAGNDATA database. Our algorithm works as follows:

When looping over all entries in the MAGNDATA database, we first check if they are collinear and then determine their \textit{bulk} spin Laue group. If it is antiferromagnetic \cite{Smejkal2021a}, then this material is a candidate for our surface analysis. (Note that this step excludes, in particular, bulk altermagnets.)
If a material has passed the previous test, we find a finite set $\mathcal{N}$ of candidate surface normals $\boldsymbol{n}$ that are left invariant under a subset of symmetries of the material's space group. 

For each surface normal $\boldsymbol{n} \in \mathcal{N}$, we calculate the nontrivial sSSG $R_{\boldsymbol{n}}^s$ and identify the corresponding nontrivial sSPG $\mathcal{R}_{\boldsymbol{n}}^s$. 
Next, we calculate the  surface spin Laue group $\mathbb{Z}^{[E||\mathcal{P}]}\times\mathcal{R}_{\boldsymbol{n}}^s$, which takes into account the effective inversion symmetry of the spin-only group for collinear magnets. At this point, we are able to identify from the surface spin Laue group if the surface is ferromagnetic (magnetically uncompensated), antiferromagnetic, or altermagnetic -- if the latter, we use Tab.~\ref{tab:tabsummary} to determine the symmetry type of altermagnetism. As a last step, we check for the element $[C_{2\perp}||C_{2\boldsymbol{n}}]$ in the surface spin Laue group to decide whether the group describes  SAM or SD-SAM.

\subsection{Effect of terrace domains on surface spin-splitting}
\label{sec:terraces}
\change{In this section, we discuss how surface terminations affect the spin splitting in a surface altermagnet.
Generally, the surface of a material is not atomically flat, giving rise to what is commonly referred to as surface roughness. In particular, if multiple terminations within the bulk unit cell have comparable surface energies, they may coexist across the sample.
Such surfaces can be thought of as a collection of terraces, each corresponding to a distinct termination. The height difference between terraces is determined by the relative positions of these terminations within the bulk unit cell. It is important to understand the impact of these terraces on the spin splitting, to assess the robustness of surface altermagnetism on real materials.}

\change{In the following, we address the symmetry conditions under which different terraces in a surface altermagnet can exhibit opposite sign of the spin splitting. In such a situation, experimental probes that effectively average over multiple terraces, like spin-resolved ARPES, would observe a diminished signal, even when the sample consists of a single magnetic domain. We will refer to this situation as the presence of \textit{terrace domains}.}

\change{The key requirement is the presence of a bulk spin space group symmetry that includes a translational component parallel to the surface normal. This symmetry must relate distinct surface terminations in such a way that their corresponding spin-splitting is reversed.}
\change{
Relating terraces with opposite spin-splitting requires identifying symmetry operations connecting them that reverse the spin expectation value. Besides the nonsymmorphic time reversal $\mathcal{T}\boldsymbol{t}$, such a mapping could arise if a surface spin space group element $[C_2 \parallel R | E]$ is accompanied by a bulk non-symmorphic element $[E \parallel R^{-1}|\boldsymbol{t}]$ with $\boldsymbol{t}\cdot\boldsymbol{n} \neq 0$. The combined symmetry $[C_2 \parallel E | R\boldsymbol{t}]$ would enforce opposite spin-splitting for the symmetry-connected terraces. For collinear magnets, however, this will always imply the presence of a nonsymmorpihc time reversal symmetry operation. Therefore, the presence of $\mathcal{T}\boldsymbol{t}$ (with $\boldsymbol{t}\cdot\boldsymbol{n} \neq 0$) in the bulk spin space group provides a complete criterion for determining whether a surface altermagnet is susceptible to terrace domains. We point out that the surface spin symmetries themselves do not depend on the surface termination.}

\change{We can determine whether a SAM or SD-SAM has terrace domains solely based on the symmetries of the bulk antiferromagnet. In particular, we can distinguish two cases: (i) Systems with $\mathcal{T}\boldsymbol{t}$, and (ii) systems without $\mathcal{T}\boldsymbol{t}$, in which Kramers degeneracy is enforced by $\mathcal{PT}$ symmetry. In case (i), surface altermagnetism emerges from breaking $\mathcal{T}\boldsymbol{t}$, since the translation is required to have  a component parallel to the surface normal. As a result, terraces connected by $\mathcal{T}\boldsymbol{t}$ will exhibit opposite spin splitting, leading to the partial cancellation of time-reversal odd signals at the surface.}
\change{In contrast, in case (ii) there is no symmetry operation relating opposite spin polarization across terraces. For terraces that are not connected by any symmetry, the magnitude of the spin splitting might depend on the termination. However, the partial-wave character ($d$, $g$ or $i$) of the spin splitting remains invariant among these terraces. Since there is no symmetry enforcing cancellation, one should expect measurable spin splitting and other time-reversal odd signals at the surface, even in presence of roughness.} \change{In Sec. \ref{Material candidates} we show a terrace domain analysis on a realistic material candidate.}

\change{We thus conclude that the occurrence of terraces domains in surface altermagnets can be predicted directly from bulk symmetries: systems possessing a $\mathcal{T}\boldsymbol{t}$ symmetry necessarily exhibit terraces domains, whereas systems lacking $\mathcal{T}\boldsymbol{t}$ but preserving $\mathcal{PT}$ symmetry do not.}

\begin{figure}\label{pie_chart}
\includegraphics[width=\linewidth]{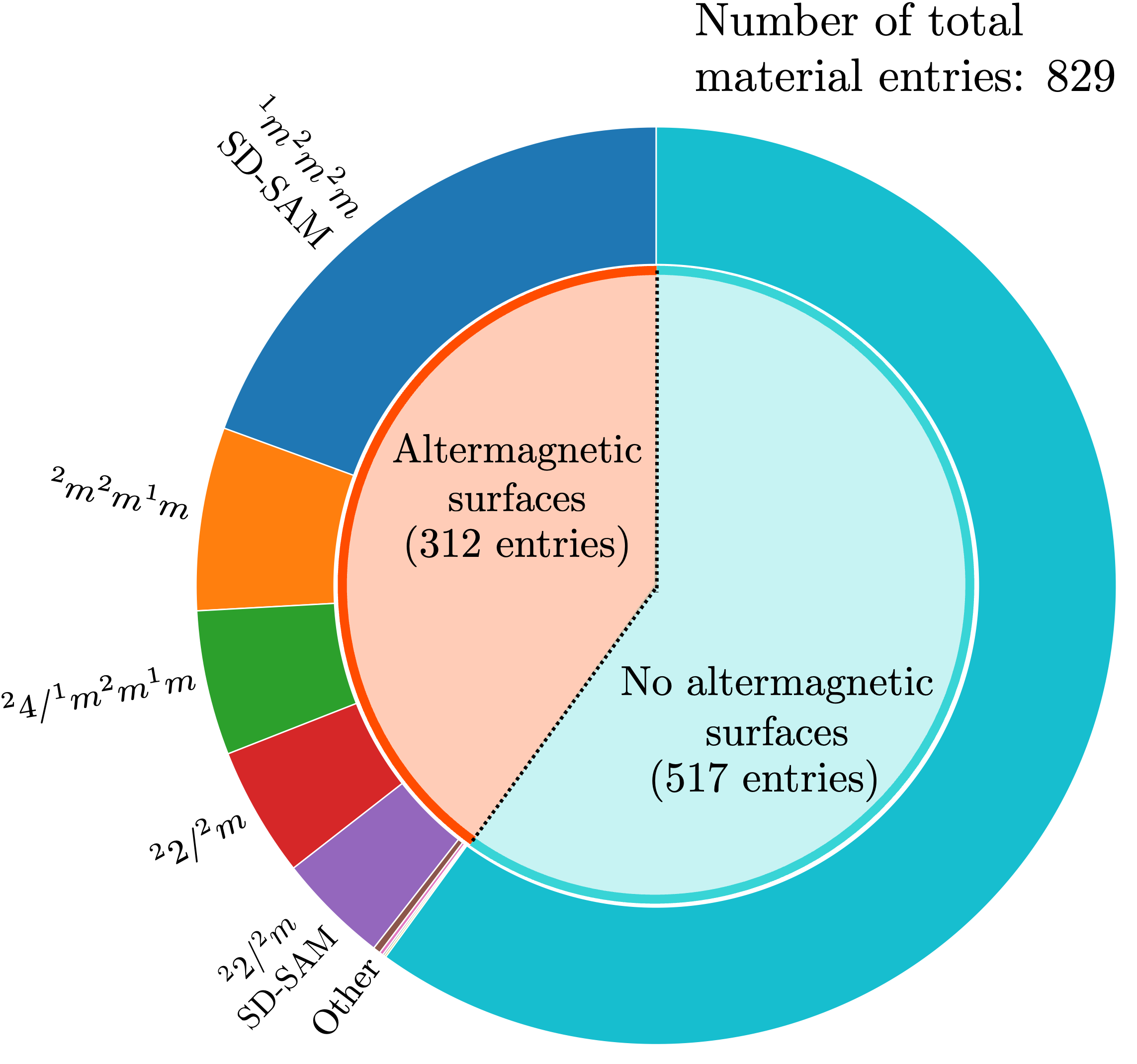}
\caption{
Pie chart summarizing the results of the surface altermagnetism screening of the MAGNDATA database, consisting of \change{829} total entries that are 3D AFMs. 
The inner pie chart shows the share of materials that have altermagnetic surfaces, i.e., SAM or SD-SAM surfaces. The outer pie chart shows the shares of the respective surface spin Laue groups. Notice that it is common among the materials with altermagnetic surfaces to have multiple different surfaces with SAM/SD-SAM 
surface spin Laue groups. The ``Other'' category encompasses the 
surface spin Laue groups $^1\bar{3}^2m$, $^26/^2m$, $^14/^1m^2m^2m$, and $^24/^1m$. The total number of surfaces for the respective surface spin Laue group is given in Tab.~\ref{tab:tabsummary}.} 
\label{fig:piechart}
\end{figure}

\subsection{List of surface altermagnet candidates in MAGNDATA database} 
\label{sec:screening-results}

The results of the screening of the MAGNDATA database \cite{Gallego2016a} are summarized in Fig.~\ref{fig:piechart}.
We identified \change{829} materials entries in the MAGNDATA database as collinear 3D AFMs. This step was done by direct determination of the spin space groups \cite{Shinohara2024} of the magnetic structures reported by neutron scattering experiments. 
While all materials trivially have some uncompensated, i.e., ferromagnetic, surfaces, we do not discuss these here, since our interest lies in compensated magnetic surfaces.
Of those with compensated surfaces, we find \change{154} materials entries that exhibit at least one SAM surface and \change{202} materials entries that exhibit at least one SD-SAM surface. There are many materials that have multiple compensated surfaces and show different sSSG that are SAM, SD-SAM, or antiferromagnetic in the same material. 
\change{Based on the conditions of having relevant terrace domains, we classify all entries in our material table and identify 161 materials without terrace-domain-cancellation effects (SAMs and SD-SAMs), as well as 151 materials in which terrace domains with opposite spin splitting are allowed. This indicates that the absence of an atomically flat sample is not generally a limiting factor for the experimental realization of surface altermagnetism.}
For each of the 11 surface spin Laue groups, we indicate in the last column of Tab.~\ref{tab:tabsummary} how many surfaces of this type in the MAGNDATA materials entries we have found, with $d$-wave SAM and SD-SAM being the dominating type of surface altermagnetism. The detailed list of MAGNDATA materials entries with altermagnetic surfaces is shown in Appendix~\ref{fulllist} in Tab.~\ref{tab:table}, including NaMnP, \change{KV$_2$Se$_2$O} and FeGe$_2$, the \change{three} compounds we address by first-principles calculations below in Sec.~\ref{Material candidates}.

\begin{figure*}[t!]
  \centering
     \includegraphics[width=\linewidth]{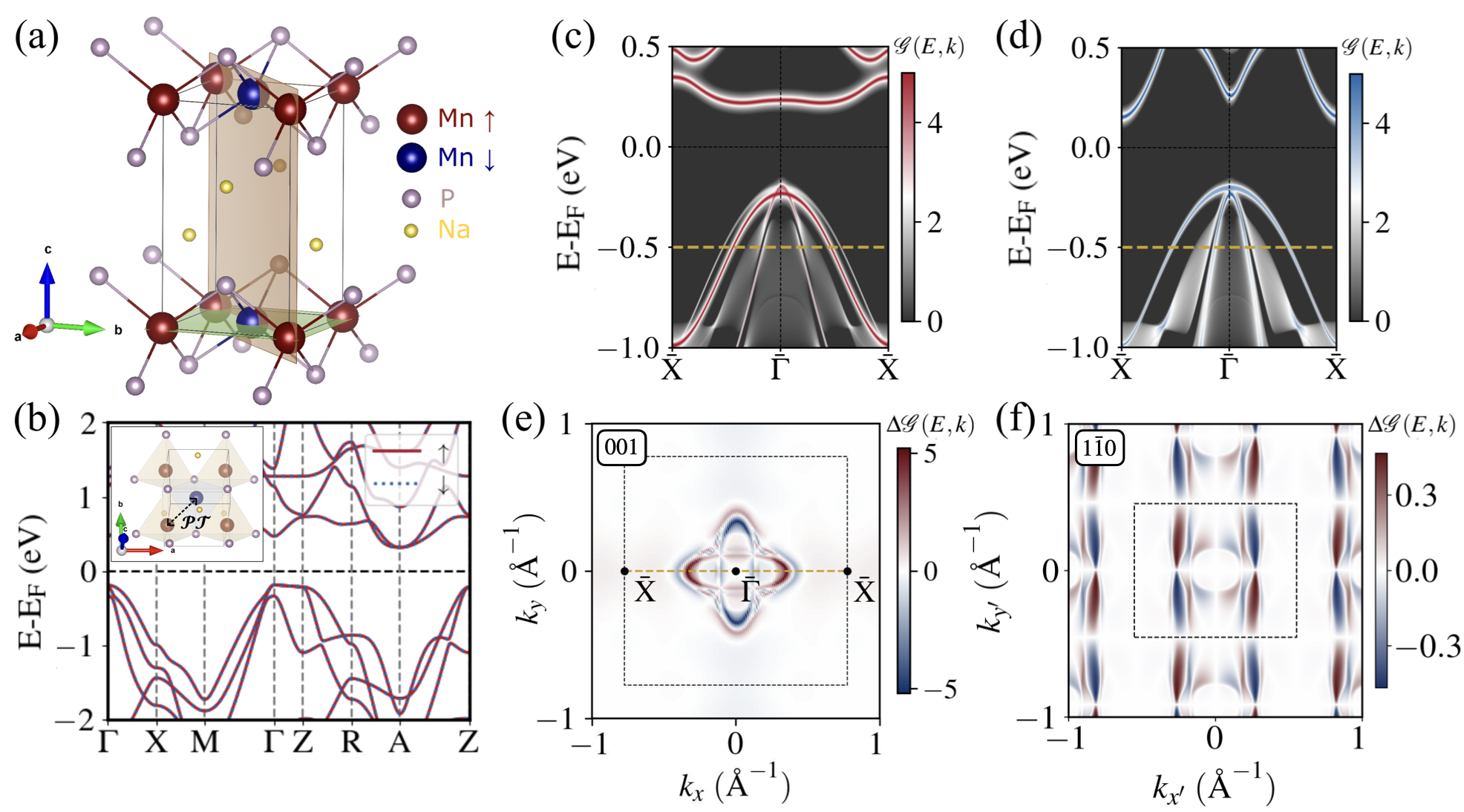}
    \caption{Candidate material NaMnP for a $\mathcal{PT}$ AFM that exhibits $d$-wave altermagnetic surface states. (a) Crystal unit cell showing compensated collinear magnetic ordering of Mn moments. Two different termination planes (001) and (1$\bar{1}$0) are indicated in green and brown. Surface spin group analysis predicts $d$-wave altermagnetic surface states for (001) and (1$\bar{1}$0), respectively, but with different surface spin Laue groups. (b) Spin-polarized bulk electronic bands calculated without spin-orbit coupling (SOC), showing Kramers' spin degeneracy. Inset highlights how the $\mathcal{PT}$ symmetry connects the two sublattices in the bulk.
    (c) Spin-up and (d) spin-down channel spectral function ($\mathcal{G}(E,\mathbf{k})$) without SOC of the (001) Mn-terminated surface, indicated with the green plane in (a), showing both bulk and surface states along the path $\bar{\mathrm{\Gamma}}$ (0.0,0.0)-$\bar{\mathrm{X}}$ (0.5,0.0), indicated by a yellow line in panel (e) in the the surface Brillouin zone.
    (e) The spin-polarized spectral function difference ($\Delta \mathcal{G}(E,\mathbf{k})=\mathcal{G}_{\uparrow}(E,\mathbf{k})-\mathcal{G}_{\downarrow}(E,\mathbf{k})$) at $E_{F} - 0.5$ eV for (001) termination exhibiting $d$-wave nature with two nodal planes, and rotational symmetry. Surface Brillouin zone indicated with a black dashed square. (f) The spectral function difference for the two spin channels calculated for (1$\bar{1}$0) termination at $E_{F} - 0.5$ eV, with the axes system rotated such that $k_{z'}$ parallel to the surface normal,  exhibiting $d$-wave nature with two nodal surfaces without rotational symmetry, and with surface Brillouin zone indicated with black dashed rectangle.}
    \label{fig:NaMnP}
\end{figure*}
 
Given the extent of the materials list provided in Tab.~\ref{tab:table} in Appendix~\ref{fulllist}, it is worthwhile to comment on certain characteristic materials. 
Interestingly, the collinear tetragonal phase of Mn$_3$Pt \cite{Kren1966} -- a well known non-collinear AFM at lower temperatures -- supports $d$-wave character SAM on its $[001]$ surfaces.
Within the screened database, only SrRu$_2$O$_6$ \cite{Tian2015b} supports $i$-wave SAM at its $[001]$ surface, and only the $[001]$ surface of KTb$_3$F$_{12}$ \cite{Guillot2002} has the surface spin Laue group $^24/^1m$ with generally curved nodal paths. 
Other materials of note are, for example, the magnetoelectric Cr$_2$O$_3$ \cite{Fiebig1996}, and the prototypical materials from antiferromagnetic spintronics, Mn$_2$Au \cite{Barthem2013} and CuMnAs \cite{Karigerasi2022}, where
$\text{Mn}_2\text{Au}$ is a prominent example of the SD-SAM case with four surfaces of this type. 

There are multiple superconducting parent compounds with altermagnetic surfaces, such as the $(010)$ surface of $\text{PrFeAsO}$ \cite{Kimber2008}, which exhibits $d$-wave SAM.
Also, $\text{La}_2\text{CuO}_4$  \cite{Reehuis2006} in its AFM phase exhibits SD-SAM in its $(010)$ surface -- note that for
$\text{La}_2\text{CuO}_4$ an AM phase has also been reported \cite{Lane2018}.
These compounds are particularly interesting from the perspective of interface and proximity effects, some of which have been studied in the context of the interplay between superconductivity and altermagnetism for generating spin-triplet Cooper pairs \cite{Chakraborty2024a,Fukaya2025}.

\section{Characteristic material candidates}
\label{Material candidates}

From the materials listed in Appendix \ref{fulllist}, we focus on three representative examples to demonstrate how surface altermagnetism is borne out in first-principles calculations. Specifically, we consider the $\mathcal{PT}$-symmetric antiferromagnets NaMnP~\cite{Bronger1986} (MAGNDATA entries No.~0.626-0.628), FeGe$_2$~\cite{Murthy1965} (MAGNDATA entry No.~1.557) \change{and KV$_2$Se$_2$O~\cite{Sun2025a} (MAGNDATA entry No.~1.875)}. NaMnP exhibits a $d$-wave surface spin polarization, whereas FeGe$_2$ presents an example of $g$-wave texture in the surface spin splitting. Finally KV$_2$Se$_2$ exhibits $d$-wave surface altermagnets as indicated by the ARPES studies in Ref. \cite{Jiang2025}.
The N\'{e}el temperatures of the antiferromagnetic NaMnP, FeGe$_2$, \change{and KV$_2$Se$_2$ are reported to be $>$293~K~\cite{Bronger1986}, $\sim$315~K~\cite{Murthy1965}, and 400~K \cite{Sun2025a}, respectively.} 
Below, we discuss the bulk and surface characteristics for these two individual materials candidates from an {\it ab initio} perspective. The specific computational details of the first principles calculations can be found in Appendix~\ref{comp-details}.

\begin{figure*}[t!]
  \centering
     \includegraphics[width=\linewidth]{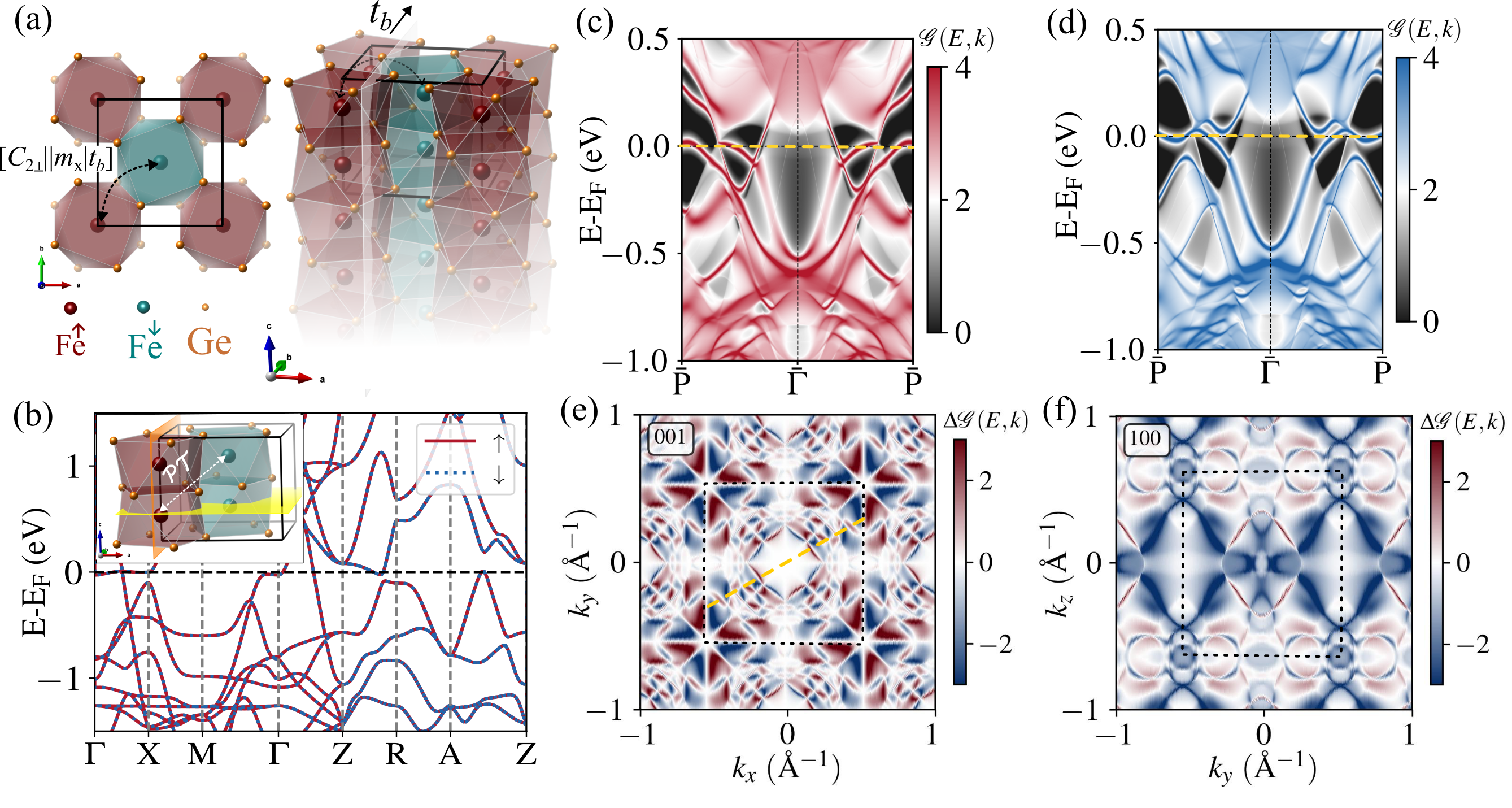}
    \caption{Candidate material $\text{FeGe}_2$ for a $\mathcal{PT}$ AFM that exhibits $g$-wave altermagnetic surface states. (a) Left panel: Top view of the crystal unit cell showing antiparallel magnetic ordering of Fe moments, indicating the [$C_{2\perp}||m_x|t_b$] spin transposing symmetry, where $t_b=(0,\frac{1}{2},0)$.   Right panel: Terminated bulk structure, showing the glide $m_x$ plane.
    (b) Spin-polarized Kramers' degenerate bulk dispersion calculated without spin-orbit coupling (SOC). The $\mathcal{PT}$ symmetric magnetic sublattices are shown in the inset. (c,d) Spectral function ($\mathcal{G}(E,\mathbf{k})$) of the (001) Fe-terminated surface [shown with yellow plane in the inset of (b)] without SOC, showing both bulk and surface states for spin-up (c) and spin-down (d) channels, respectively. The spin splitting near the Fermi energy is $\sim$0.08 eV along the path $\bar{\mathrm{\Gamma}}$ (0.0,0.0)-$\bar{\mathrm{P}}$ (0.5,0.25), indicated by a dashed yellow line in panel (e) in the surface Brillouin zone (dashed black square). (e) The spin-polarized spectral function difference ($\Delta \mathcal{G}(E,\mathbf{k})=\mathcal{G}_{\uparrow}(E,\mathbf{k})-\mathcal{G}_{\downarrow}(E,\mathbf{k})$) at the Fermi energy for (001) termination exhibits $g$-wave nature with four nodal planes. (f) The spectral function difference for two spin channels calculated for (100) termination [shown with orange plane in the inset of (b)], indicating an uncompensated ferromagnetic nature in agreement with symmetry prediction. 
    }
    \label{fig:FeGe2}
\end{figure*}

\subsection{$D$-wave surface altermagnetism in NaMnP}
\label{sec:NaMnP}

NaMnP is a bulk antiferromagnetic semiconductor with collinear compensated magnetic moments on the Mn atoms \cite{Bronger1986, Chen2025b}, see Fig.~\ref{fig:NaMnP}(a). This ordering has a magnetic point group $4'/m'm'm$. NaMnP crystallizes 
with space group $P4/nmm$, where the Mn atoms sit on Wyckoff position 2a~$(0,0,0)$, and the Na and P on position 2c~$(\frac{1}{2}, 0, z)$. The crystal structure is closely related to the previously discussed Lieb lattice. However, in this case, the nonmagnetic ligands \textit{do not} break the center of inversion that connects the two oppositely spin-polarized Mn sublattices,  as illustrated by the inset in Fig.~\ref{fig:NaMnP}(b). As a result, the crystal structure more closely resembles the symmetry of FeSe \cite{Mazin2023a}. The orientation of the ligand pyramids in NaMnP breaks $\mathcal{T}\mathbfit{t}$ symmetry, making NaMnP a bulk $\mathcal{PT}$ antiferromagnet. As a consequence, Kramers' spin degeneracy is preserved for the bulk bands everywhere in momentum space, as can be seen in the spin-polarized band structure we computed in the non-relativistic limit in Fig.~\ref{fig:NaMnP}(b). Our {\it ab initio} calculations yield a magnetic moment of 3.51~$\mu_B$ per Mn site, in good agreement with the experimentally reported moment of magnitude 3.64~$\mu_B$~\cite{Bronger1986}.

Clearly, the symmetry that preserves the effective TRS is broken when terminating the material in the $[001]$ or in the $[1\bar{1}0]$ (and $[110]$) directions. However, the effect of the surface termination on the other symmetries differs between the two surfaces. 
The surface with normal $\hat{\boldsymbol{n}}\parallel [001]$, preserves some of the rotational symmetries of the bulk, including the $[C_{2\perp}||C_{4z}]$ symmetry, resulting in the surface spin Laue group $^{2}4/^{1}m ^{1}m ^{2}m $. 
We demonstrate the distinct nature of the surface by studying the spectral function $\mathcal{G}_{\sigma}(E,\boldsymbol{k})$ of the spin-up ($\sigma = \uparrow$) and spin-down ($\sigma = \downarrow$) polarized bands at the surface. It is given by $\mathcal{G}_{\sigma}(E,\boldsymbol{k}) = - \text{Im} \text{Tr} G_{\sigma}(E,\boldsymbol{k}) /\pi$, where $G_{\sigma}(E,\boldsymbol{k})$ is the surface Green's function of the respective spin channel $\sigma$. For the surface with normal $\hat{\boldsymbol{n}}\parallel[001]$, we choose the Mn atom termination and show $\mathcal{G}_{\sigma}(E,\boldsymbol{k})$ in Fig.~\ref{fig:NaMnP}(c,d), for the spin-up and spin-down channels, respectively. Here, the white shaded region predominantly captures the bulk contribution to the spectral functions, which is equal in both channels, while the red and blue regions indicate the surface states, which are distinct. Furthermore, in Fig.~\ref{fig:NaMnP}(e), we show the difference in the spectral function between the two polarizations, i.e., $\Delta \mathcal{G}(E,\boldsymbol{k})=\mathcal{G}_{\uparrow}(E,\boldsymbol{k})-\mathcal{G}_{\downarrow}(E,\boldsymbol{k})$, for the $(001)$ surface, again with the surface termination of Mn atoms. This difference spectral function shows the strong polarization with the expected $d$-wave nature, with two nodal lines, which are protected by the spin symmetries $[C_{2\perp}||m_{[110]}]$ and $[C_{2\perp}||m_{[1\bar{1}0]}]$. The difference spectral function changes sign under a four-fold rotation, as expected from the symmetry $[C_{2\perp}||C_{4z}]$ of the 2D surface. We note that for a surface with the same normal, but termination with the nonmagnetic P and Na atoms, we find a similar effect with identical symmetry. The surface with normal $\hat{\boldsymbol{n}}\parallel[1\bar{1}0]$ only preserves two mirrors, and breaks the four-fold rotational symmetry, resulting in the surface spin Laue group $^{2}m ^{2}m ^{1}m$ and a different $d$-wave character than the $(001)$ termination. We show the difference spectral function for this surface termination in Fig.~\ref{fig:NaMnP}(f), with the axes rotated such that $z'$ indicates the surface normal (and $k_{x'}$ and $k_{y'}$ span the surface). Again, we see the clear $d$-wave nature, with two nodal planes which are protected by the spin symmetries $[C_{2\perp}||m_{x'}]$ and $[C_{2\perp}||m_{y'}]$, but here without the four-fold rotational symmetry.

\subsection{$G$-wave surface altermagnetism in FeGe$_2$}
\label{sec:FeGe2}

\begin{figure*}[t!]
  \centering
     \includegraphics[width=\linewidth]{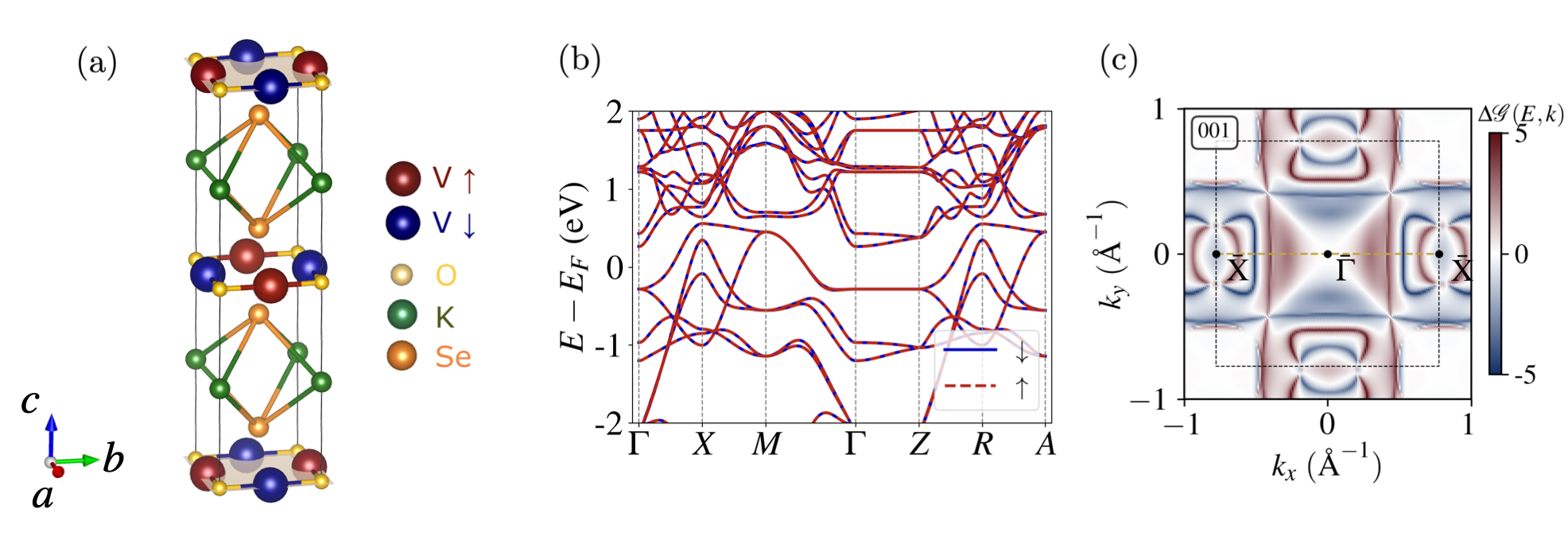}
    \caption{
    \change{Surface altermagnetism in KV$_2$Se$_2$O. (a) Magnetic unit cell of the AFM order~\cite{Sun2025a}. The surface termination is highlighted in brown on the VO planes. (b) Electronic band structure without spin orbit coupling showing Kramers degeneracy. (c) Spin-resolved surface spectral function $\Delta\mathcal{G}(E,\mathbf{k})$ at the Fermi energy, calculated for the termination indicated in panel (a). The spin-splitting texture exhibits $d$-wave symmetry.}
    }
    \label{fig:kv2se2O}
\end{figure*}

The second candidate material FeGe$_2$ [see Fig.~\ref{fig:FeGe2} (a)] provides a representative example for the realization of $g$-wave surface altermagnetism. It crystallizes in the tetragonal $I4/mcm$ space group~\cite{Murthy1965}, with Fe and Ge atoms occupying the Wyckoff positions 4a~$(0,0,\frac{1}{2})$ and 8h~$(x,x+\frac{1}{2}, \frac{1}{4})$, respectively. The bulk magnetic ground state is a collinear antiferromagnetic configuration belonging to the magnetic point group $mmm1'$, which preserves $\mathcal{PT}$ symmetry. We have computed the spin-polarized band structure 
in the non-relativistic limit in Fig.~\ref{fig:FeGe2}(b). 
Our {\it ab initio} GGA calculations yield a magnetic moment of 1.53~$\mu_B$ per Fe site, in good agreement with the experimentally reported effective moment of 1.21~$\mu_B$~\cite{Murthy1965}. 
Consistent with bulk $\mathcal{PT}$ symmetry, Kramers' degeneracy is preserved across the entire Brillouin zone, as evident from the bulk band structure shown in Fig.~\ref{fig:FeGe2}(b). The $\mathcal{PT}$ partners of magnetic Fe sublattices are shown in the inset of Fig.~\ref{fig:FeGe2}(b). 

Applying symmetry arguments analogous to those previously discussed for the $d$-wave surface altermagnet NaMnP, we predict that FeGe$_2$ supports a $g$-wave altermagnetic surface state for a $(001)$ surface termination. {The bulk magnetic structure has the spin symmetry $[C_{2\perp}||m_x|\mathbfit{t}_b]$, with $\mathbfit{t}_b=[0,\frac{1}{2},0]$, which connects opposite-spin Fe sublattices, as shown in Fig.~\ref{fig:FeGe2}(a). This glide mirror, together with three additional glide mirrors normal to $[110]$, $[1\bar{1}0]$ and $[010]$, is preserved under the $(001)$ surface termination, and lead to the surface spin Laue group $^14/^1m^2m^2m$}.
The material-specific schematic of the semi-infinite geometry is depicted in the right panel of Fig.~\ref{fig:FeGe2}(a). 
We again have calculated the spectral functions $\mathcal{G}(E,\boldsymbol{k})$ at the $\hat{\boldsymbol{n}}\parallel[001]$ Fe terminated surface in Fig.~\ref{fig:FeGe2}(c,d) for spin-up and spin-down channels, respectively, in the non-relativistic limit. While the bulk spectral function (white shaded region) remains identical for both spin channels, the surface states (red and blue) exhibit spin-dependent momentum dispersion. 
As with NaMnP, to characterize the symmetry of the surface spin-polarization, we show the difference of the spin-resolved spectral functions,  $\Delta \mathcal{G}(E,\boldsymbol{k})=\mathcal{G}_{\uparrow}(E,\boldsymbol{k})-\mathcal{G}_{\downarrow}(E,\boldsymbol{k})$, in Fig.~\ref{fig:FeGe2}(e) for the (001) termination. The resulting pattern reveals a clear $g$-wave texture in the 2D surface Brillouin zone with nodal lines coinciding with mirrors $m_x$, $m_y$, $m_{110}$, and $m_{1\overline{1}0}$ contained within the surface spin Laue group $^14/^1m^2m^2m$.
For another choice of consecutive Fe termination, the $g$-wave texture remains the same but reverses its sign. This sign reversal of the texture can be understood from the fact that the local environments of two successive Fe layers with opposite spins are related by a $[C_{2\perp}||C_{2y}|\boldsymbol{t}_b]$ symmetry (two-fold rotation in spin-space, combined with a two-fold screw axis in real space). Notably, the altermagnetic $g$-wave texture of $\Delta \mathcal{G}(E,\boldsymbol{k})$ is robust against different surface terminations, whether by magnetic Fe layer or nonmagnetic Ge layer. 

To further validate our symmetry-based analysis in a realistic setting, we have calculated $\Delta \mathcal{G}(E,\boldsymbol{k})$ for $\hat{\boldsymbol{n}}\parallel[100]$ which does \textit{not} support surface altermagnetism. The corresponding surface is shown in the inset of Fig.~\ref{fig:FeGe2}(b) by the orange plane. In this case, $\Delta \mathcal{G}(E,\boldsymbol{k})$ exhibits uncompensated or ferromagnetic character [see Fig.~\ref{fig:FeGe2}(f)] with one spin channel dominating the spectral weight for an Fe terminated $\hat{\boldsymbol{n}}\parallel[100]$ surface, which is also in agreement with our symmetry based analysis.

\subsection{$D$-wave surface altermagnet KV$_2$Se$_2$O}
\label{sec:KVSeO}
\change{The emergent altermagnetism at the surfaces of antiferromagnets discussed above provides a natural route toward reconciling the apparently conflicting experimental reports on KV$_2$Se$_2$O. The angle-resolved photoemission spectroscopy (ARPES) data of Ref.~\cite{Jiang2025} were interpreted in terms of \emph{bulk} $d$-wave altermagnetism. However, this interpretation appears to be at odds with the neutron diffraction results of Ref.~\cite{Sun2025a}, which instead indicate a $G$-type antiferromagnetic order in the bulk, characterized by a $(0,0,\frac{1}{2})$ propagation vector, as indicated in Fig.~\ref{fig:kv2se2O}(a). To demonstrate how the concept of surface altermagnetism resolves this controversy, we present DFT calculations for the (001) surface, which was measured in the ARPES study of Ref.~\cite{Jiang2025}.}
\change{Our calculations were done in the nonrelativistic limit, using the double AFM unit cell of KV$_2$Se$_2$O. We obtain a magnetic moment of 1.67 $\mu_B$ per V site, reasonably close to the value 1.31 $\mu_B$, reported by the neutron diffraction experiment \cite{Sun2025a}. The electronic band structure is spin degenerate in the bulk, as shown in Fig.~\ref{fig:kv2se2O}(b), due to the $\mathcal{T}\boldsymbol{t}$ symmetry introduced by the nonzero propagation vector.}

\change{The bulk spin point group of the material is given by $\mathcal{R}_s = \mathbb{Z}_2^{[C_{2\perp}||E]}\times {}^14/{}^1m{}^1m{}^1m$, as in the minimal model presented in Sec.~\ref{Toy model}. 
When looking at the surface normal to $(001)$, the resulting surface spin point group is given by ${}^24{}^2m{}^1m$, which classifies the surface as altermagnetic. The generators of this group are therefore $[C_{2\perp}||C_4]$, $[C_{2\perp}||m_{110}]$, and $[E||m_{001}]$.}

\change{Because ARPES is a surface-sensitive probe that predominantly detects electronic states in proximity to the surface termination of the system, the surface spin point group is the appropriate object for predicting the spin-resolved spectra. Therefore, our results resolve the apparent discrepancy: ARPES is inherently surface sensitive, and the $(001)$ surface of the proposed AFM order in KV$_2$Se$_2$O reduces the symmetry, allowing for a SAM characterized by the $d$-wave surface spin Laue group ${}^24/{}^1m{}^1m{}^2m$ (which is equivalent to the ${}^24/{}^1m{}^2m{}^1m$ group reported in Tab.~\ref{tab:table}, but with a rotated $d$-wave pattern). This is further confirmed by our \textit{ab initio} calculations: the electronic bulk dispersion is spin-degenerate, as shown in  Fig.~\ref{fig:kv2se2O}(b), but the spin-resolved surface spectral function shown in Fig.~\ref{fig:kv2se2O}(c) [calculated for the termination indicated in Fig.~\ref{fig:kv2se2O}(a)] exhibits a clear $d$-wave pattern. These calculations establish KV$_2$Se$_2$O as a strong candidate for surface altermagnetism, in agreement with both ARPES and neutron diffraction studies.}

\subsection{Spin-textures for different surface terminations in material candidates}

\begin{figure*}[t!]
  \centering
     \includegraphics[width=\linewidth]{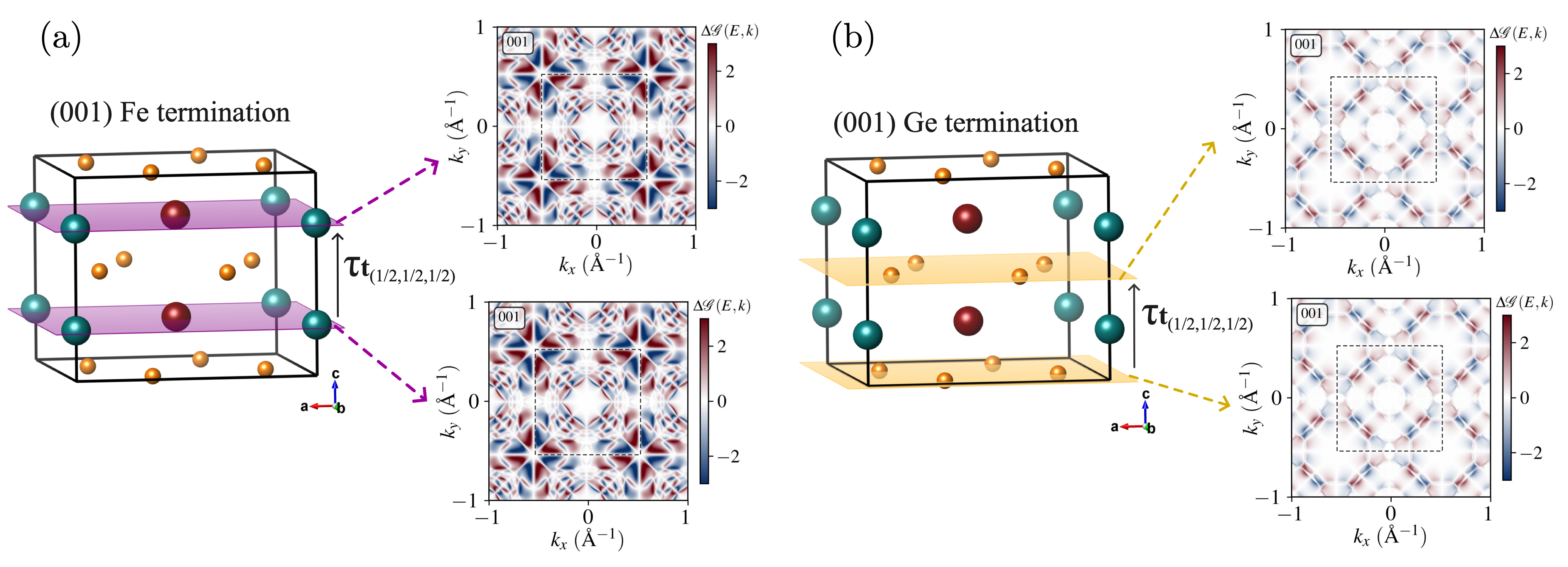}
    \caption{\change{Spin-resolved surface spectral density at the Fermi energy for different (001) surface terminations of $\text{FeGe}_2$, together with the bulk unit cell indicating the corresponding surface terminations by magenta and yellow planes. (a) Magnetic surface terminations. (b) Non-magnetic surface terminations.}
    }
    \label{fig:different_terminations}
\end{figure*}

\change{We next analyze the spin textures of different surface terminations in the aforementioned compounds.} 

\change{
In Fig.~\ref{fig:different_terminations}, we show the unit cell of the candidate material for $g$-wave surface altermagnetism, $\text{FeGe}_2$, together with four surface terminations and the corresponding spin-resolved surface spectral functions, evaluated at the Fermi level.
}
\change{As discussed in Sec.~\ref{sec:FeGe2}, $\text{FeGe}_2$ has a $\mathcal{T}\boldsymbol{t}$ symmetry in the bulk. The surface terminations in Fig.~\ref{fig:different_terminations}(a) and Fig.~\ref{fig:different_terminations}(b) with surface normal $(001)$, taken in magnetic and non-magnetic layers, respectively, are related by this $\mathcal{T}\boldsymbol{t}$ symmetry. As a result, the spectral functions are symmetry-enforced to have identical magnitude and opposite sign. Thus, the spin splitting, and other time reversal-odd signals are expected to vanish for equally distributed terraces, as contributions from $\mathcal{T}\boldsymbol{t}$-related terminations cancel each other.}

\change{We further highlight the differences in the magnitude of the spin splitting for different surface terminations. One observes that terminations containing magnetic atoms [see Fig.~\ref{fig:different_terminations}(a)] exhibit a larger overall spin splitting than those terminated by non-magnetic atoms [see Fig.~\ref{fig:different_terminations}(b)]. This observation can be explained by the orbital characters of the surface states at the Fermi level. For terminations with magnetic atoms, the relevant electronic states originate from partially filled orbitals, leading to a larger spin polarization. In contrast, for non-magnetic terminations, the spin polarization arises mainly from hybridization with orbitals from adjacent magnetic atoms and is therefore reduced compared to the magnetic-atom termination.}

\change{Such terrace domains are also symmetry allowed in KV$_2$Se$_2$O, in which experiments clearly observe altermagnetic signatures in the ARPES spectra while simultaneously measuring a magnetic order that would enforce spin degenerate bulk band structure. Terrace domains are therefore not necessarily detrimental to the experimental realization of surface altermagnetism in a given material.}

\change{No terrace domains are expected in NaMnP.}

\section{Discussion}
\label{discussion}


Surface altermagnetism naturally opens up perspectives for interface and device physics. In particular, one may envision insulating three-dimensional antiferromagnets whose altermagnetic surface or interface states are metallic, forming a two-dimensional electron gas that is localized near the boundary. Since bulk charge transport is suppressed in such systems, the altermagnetic properties of the surface or interface would be directly accessible in transport measurements. For instance, a two-dimensional electron gas with $d$-wave altermagnetic symmetry is expected to exhibit the spin-splitter effect. More generally, the theoretical framework of surface spin groups introduced here can be directly extended to \emph{interface spin groups}, allowing one to systematically describe the combined spin symmetries of the interfaced materials together with the symmetry reduction intrinsic to the interface.


\change{We note that the concept of surface spin groups used in this work has also appeared in Ref.~\cite{Zhou2025}, building on the earlier theory of surface magnetic groups in Ref.~\cite{Weber2024}. While Ref.~\cite{Zhou2025} focuses on free-standing slab geometries, we develop a systematic classification of surface spin groups for semi-infinite systems relevant to ARPES and other surface-sensitive probes.}

\change{Furthermore, the experimental signatures currently used to establish altermagnetism are commonly obtained from surface- or interface-sensitive probes, making surface spin groups the natural symmetry framework in these settings, rather than bulk or strictly two-dimensional spin groups. In this context the observation of $d$-wave altermagnetism by the surface-sensitive ARPES measurements reported in Ref.~\cite{Jiang2025} is fully compatible with the bulk antiferromagnetic order reported in Ref.~\cite{Sun2025a}. In addition, since the ARPES observations on Rb$_{1-\delta}$V$_2$Te$_2$O \cite{Zhang2025a} exhibit the same structure, if Rb$_{1-\delta}$V$_2$Te$_2$O turns out to be antiferromagnetic in the bulk, \change{as indicated by recent neutron diffraction results in Ref.~\cite{Xie2026},} our theory would likewise predict a SAM with $d$-wave character in the experimental orientation. Furthermore, the ARPES spectra measured on the (001) surface in Ref.~\cite{Jiang2025} clearly exhibit the $C_4$ symmetry predicted from the surface spin point group of the system.}


While SAM surfaces exhibit the characteristic beyond $s$-wave even-parity spin splittings of $d$-, $g$-, or $i$-wave symmetry, SD-SAM surfaces do not. In the nonrelativistic limit, their electronic bands remain spin-degenerate. The key distinction from antiferromagnetic surfaces is that SD-SAM surfaces lack time-reversal symmetry in their surface point group. We therefore argue that SD-SAM surfaces become particularly relevant once relativistic effects are taken into account: spin degeneracy is lifted throughout the surface Brillouin zone, except at specific high-symmetry points determined by the orientation of the Néel vector, thereby rendering time-reversal symmetry breaking explicitly observable. Two-dimensional material systems with this symmetry have recently been shown to exhibit an anomalous Hall effect \cite{Bai2025a}. SD-SAM surfaces -- which may host sizable Rashba spin–orbit coupling due to structural inversion symmetry breaking at the surface -- can thus be expected to display pronounced magneto-optical responses. In particular, the magneto-optical Kerr effect in reflection geometry is inherently surface sensitive and a natural probe of such states. In addition, in embedded interfaces with topological insulating materials, many effects would be greatly enhanced, and possible quantum anomalous Hall effect scenarios could be envisioned. 

The interplay between altermagnetism and spin-orbit coupling is expected to be equally important for SAM surfaces. A prominent illustration of how these effects can conspire to yield unexpected spin symmetry properties is provided by the bulk altermagnet MnTe. In the nonrelativistic limit, MnTe realizes a $g$-wave altermagnetic phase dictated by its spin symmetries \cite{Smejkal2021a}. Upon inclusion of spin-orbit coupling, the broken time-reversal symmetry by the altermagnetic order produces a relativistic spin polarization orthogonal to the Néel order, on one of the nodal planes of the non-relativistic $g$-wave order, that has even parity ($d$-wave) and is time-reversal odd \cite{Krempasky2024, Lee2024}. By analogy, SAM surfaces of suitable materials may undergo a similar spin-orbit-coupling-induced transmutation of their effective altermagnetic symmetry. A systematic exploration of the interplay between SAM and relativistic effects is left for future work.

\change{We reiterate that our analysis of emergent surface altermagnetism relies exclusively on the spin symmetry of the bulk parent magnetic phase. In practice, however, surfaces may exhibit additional symmetry changes depending on whether they terminate in vacuum or form part of a heterostructure. Such modifications can arise from structural relaxation, which preserves the bulk-terminated surface unit cell, structural reconstruction, which alters it, or from changes in the magnetic order driven by modified exchange fields or surface magnetic anisotropies.}

\change{Surface relaxations, as considered here, modify only the positional parameters of the relevant Wyckoff sites and therefore preserve the symmetry of the ideal surface. This is particularly intuitive for so-called normal relaxations, in which the interlayer spacing perpendicular to the surface is modified near the termination. Since such displacements occur only along the surface normal and respect the existing surface symmetries, they do not alter the symmetry classification of the system within our framework. For surface reconstruction, such a general argument is not possible.}

\change{Although reconstructions are conceivable, they are neither universal nor predictable from bulk symmetry alone; they depend sensitively on chemical bonding, termination, and preparation. In contrast, our framework identifies surface altermagnetism as a direct symmetry-based consequence of the bulk magnetic order that is independent of surface-specific chemistry or reconstruction.}


\change{While this material-specific analysis of reconstructions lies beyond the scope of the symmetry-informed materials prediction presented here, common surface reconstructions may not necessarily be detrimental to surface altermagnetism.}
\change{In Appendix \ref{SAM_and_reconstruction} we propose a simplified picture in which the reconstructured crystal motifs at the surface do not lower the symmetries beyond those in the ideal surface termination. We conclude that, under this assumption, a $N\times N$ surface reconstruction (unit cell enlargement) would preserve the altermagnetic character of the surface. Additionally, some surface reconstructions of type $N\times M$ could for example: (i) lower the symmetry of a surface altermagnet from $g$ to $d$ wave, (ii) turn a SD-SAM into a SAM, or (iii) break the magnetic compensation. Modern interface engineering techniques further enable precise control over surface termination and strain, allowing experimentalists to  preserve bulk-derived symmetries in many systems.
These considerations show that common surface reconstructions are not detrimental to our complete classification of surface altermagnetism and can even lead to interesting novel effects. This further establishes our symmetry analysis as not only predictive but also foundational for the systematic functionalization of AFMs in spintronic applications.}

\change{The study of altermagnetism is often connected to the measurement of time-reversal odd signals, which are susceptible to the presence of magnetic and structural domains.}
\change{The possibility of having relevant terrace domains adds another mechanism for canceling altermagnetic effects. Recent studies have shown, however, that altermagnetic domain sizes can be sufficiently large to be detected in transport measurements, as indicated by XMLD/XMCD measurements in MnTe \cite{Amin2024}. We emphasize, however, that the symmetry-breaking pattern at the surface can also provide novel means to switch not only the altermagnetic domain structure, but also the Néel order of the bulk. Furthermore, as discussed in Sec.~\ref{Material candidates}, the Lieb-lattice compound KV$_2$Se$_2$O exhibits surface domains relevant to surface altermagnetism and nevertheless shows a pronounced spin polarization in surface-sensitive ARPES measurements \cite{Jiang2025}. This experimental observation demonstrates that both terrace and magnetic domains are large enough to detect characteristic signatures of surface altermagnetism.}



\section{Conclusion and Outlook}
\label{Conclusion and Outlook}

We have established a general and symmetry-driven pathway to realize two-dimensional altermagnetism through the controlled termination of bulk collinear antiferromagnets. By introducing the concept of surface spin groups and, in particular, surface spin Laue groups, we developed a systematic classification algorithm that identifies surfaces of antiferromagnets capable of hosting altermagnetic symmetries -- either spin-split (SAM) or spin-degenerate (SD-SAM) -- without requiring exfoliation or external field engineering. Applied to the MAGNDATA database \cite{Gallego2016a}, this approach reveals over \change{150} materials entries with at least one altermagnetic surface with spin splitting, vastly expanding the pool of experimentally accessible \change{material candidates which realize 2D altermagnetism through their electronic states at the surface.} 

Our minimal tight-binding model, inspired by layered Lieb-lattice systems, provides an intuitive proof of concept: surface termination in an otherwise conventional AFM  can induce an even-wave spin-splitting pattern characteristic of altermagnetism. This is confirmed by {\it ab initio} calculations on two representative materials -- NaMnP ($d$-wave SAM) and FeGe$_2$ ($g$-wave SAM) -- which exhibit clear spin-resolved surface band structures with the predicted altermagnetic nodal topology.

This paradigm shift -- from searching for bulk AMs or engineering 2D AMs via exfoliation or adsorption -- to generating 2D AMs via surface termination of abundant AFMs, opens a scalable route toward experimental realization. Crucially, these surface states are directly accessible to surface-sensitive probes such as spin-polarized STM, ARPES, and magneto-optical Kerr effect, enabling unambiguous experimental verification. 

We have also argued that the ARPES~\cite{Jiang2025} and neutron scattering~\cite{Sun2025a} measurements on KV$_2$Se$_2$O already contain signatures consistent with surface-induced altermagnetism in a bulk AFM.  This raises a cautionary note:  
phenomena previously ascribed to intrinsic bulk altermagnetic order could instead originate from the surface. 

In summary, emergent surface-induced altermagnetism provides not only a theoretical framework but also a practical materials design strategy to bridge the gap between the abundance of AFMs and the scarcity of experimentally viable 2D AMs. This approach promises to accelerate the discovery and deployment of altermagnetic materials in next-generation spintronic, quantum, and topological technologies.

\begin{acknowledgments}
This work was funded by the German Research Foundation (DFG) through TRR 173-268565370 (Projects No.~A03 and B13), TRR 288-422213477 (Projects No.~A09 and B05), and Project No.~504261060 (Emmy Noether Programme). We acknowledge support by the Dynamics and Topology Center (TopDyn) funded by the State of Rhineland-Palatinate.  LS acknowledges funding from the ERC Starting Grant No. 101165122. We also acknowledge the high-performance computational facility of the supercomputer ``Mogon" at Johannes Gutenberg-Universität Mainz, Germany.
\end{acknowledgments}

\bibliographystyle{apsrev4-2}
\bibliography{references}

\onecolumngrid

\appendix

\section{Ferroic order and spin splitting in SD-SAM and SAM}
\label{spin_density_SDAM_vs_SAM}

\begin{figure*}[h!]
  \centering
     \includegraphics[width=0.7\linewidth]{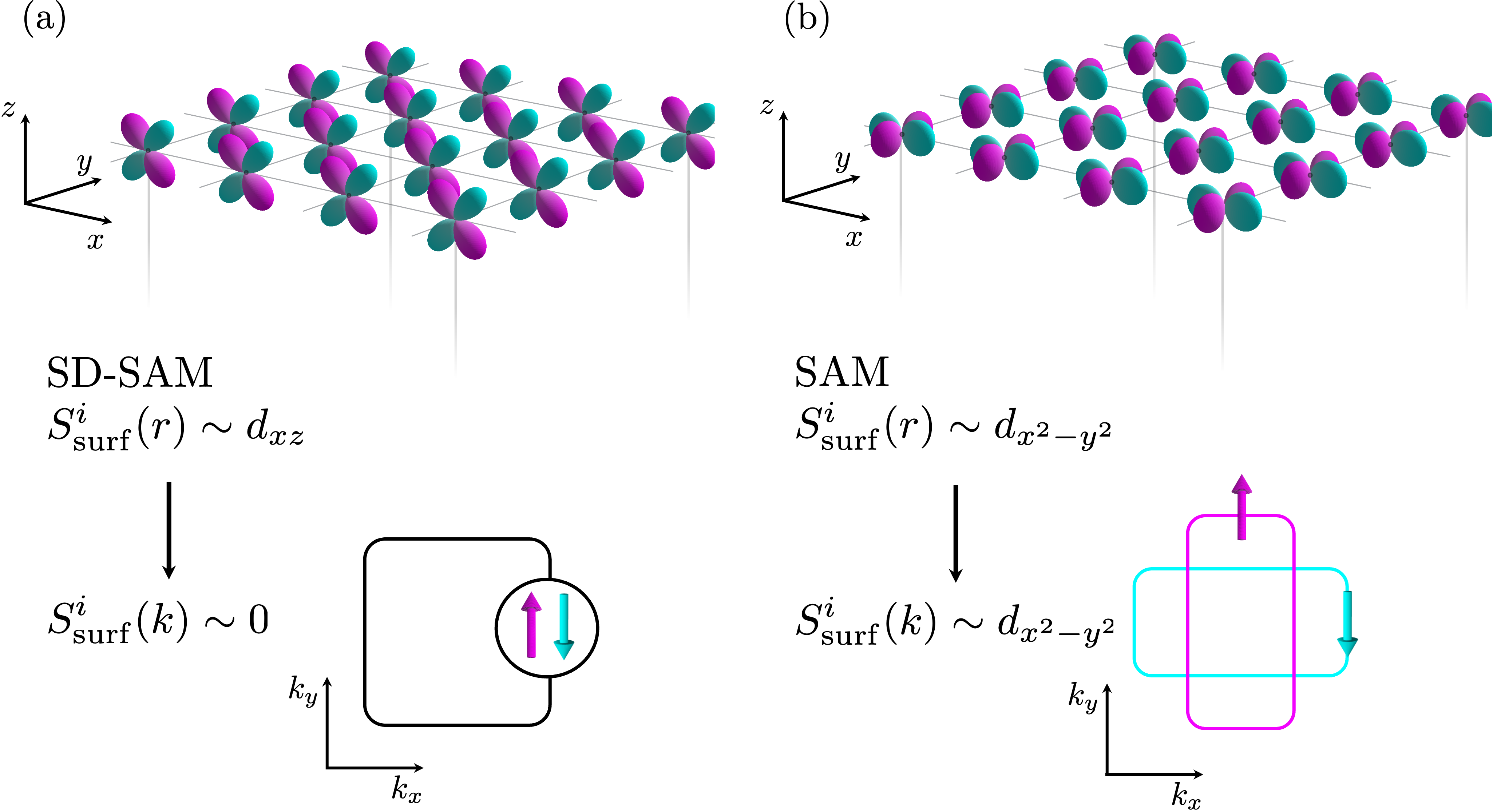}
    \caption{\change{Ferroic component of the atomic spin density in SD-SAM and SAM. (a) $d_{xz}$-shaped surface spin densities $S^i_\text{surf}(r)$ in real space with nodal plane and surface plane coinciding. This scenario  corresponds to SD-SAM and results in zero momentum-space spin polarization $S^i_\text{surf}(k)$ in the surface BZ.  (b) $d_{x^2-y^2}$-shaped surface spin densities $S^i_\text{surf}(r)$ in real space with nodal planes perpendicular to the surface plane. The surface BZ is fully spin polarized [nonzero $S^i_\text{surf}(k)$] except for the points induced by the nodal planes.}}
    \label{fig:SD_SAM_vs_SAM}
\end{figure*}

\change{In this section, we compare qualitatively the real-space order parameter defining the altermagnetism in SD-SAM and SAM. We do so by illustrating the differences in the ferroic component of their atomic spin densities (as defined in Ref.~\cite{Jaeschke-Ubiergo2025}). Figure~\ref{fig:SD_SAM_vs_SAM}(a) shows an example of an SD-SAM, in which the atomic spin density has a ferroic component described by a $d_{xz}$ spherical harmonic. This ferroic order is allowed by the symmetry $[C_2||C_{2z}]$, since a $180^\circ$ rotation about the $z$ axis transforms $xz \rightarrow -xz$, while the $C_2$ operation in spin space simultaneously reverses the sign of the collinear spin component, keeping the real-space form factor $xz$ invariant. However, because these spin-densities have a node at $z=0$, there is no spin polarization in the surface BZ.}

\change{The situation is different for the SAM scenario in Fig.~~\ref{fig:SD_SAM_vs_SAM}(b), where the atomic spin density at the surface has a ferroic component characterized by a $d_{x^2-y^2}$ spherical harmonic. This corresponds to the situation one finds, for example, at the (001) surface of KV$_2$Se$_2$O. In this case, the projection into the 2D surface Brillouin zone preserves the spin polarization and therefore gives rise to the usual altermagnetic spin-split band structure.}

\section{Computational Details}
\label{comp-details}
We performed {\it ab initio} calculations using density functional theory (DFT) within the plane-wave basis. We used the Perdew-Burke-Ernzerhof (PBE) implementation of the generalized gradient approximation (GGA) for the exchange correlation functional~\cite{Perdew1996}, together with the projector augmented-wave potentials~\cite{Blochl1994, Kresse1999} as implemented in the Vienna ab initio simulation package (VASP)~\cite{Kresse1996, Kresse1999}. All calculations were performed without Coulomb correlation and without spin-orbit coupling. In our DFT calculations we chose kinetic energy cutoffs for the plane-wave basis of 800 eV for NaMnP and KV$_2$Se$_2$O, and 550 eV for FeGe$_2$. $\Gamma$-centered $k$-point meshes of 12$\times$12$\times$7 (NaMnP), 12$\times$12$\times$3 (KV$_2$Se$_2$O) and 8$\times$8$\times$10 (FeGe$_2$) were used to perform the self cvonsistent calculations of the bulk systems.
To calculate the surface spectral functions, we constructed the Wannier functions using the VASP2WANNIER90~\cite{Marzari1997} and WANNIER90 codes~\cite{Mostofi2014}. The low-energy tight-binding Hamiltonian was defined in an effective Wannier basis including only Na-s, Mn-d, and P-p for NaMnP, V-d, Se-p, K-p and O-p for KV$_2$Se$_2$O, and Fe-d and Ge-p for FeGe$_2$, with all remaining degrees of freedom down-folded. Using the obtained tight-binding model, we calculate the surface spectral function for different terminations using the iterative Green’s function method, as implemented in the WannierTools package~\cite {Wu2017b}.

\section{Surface altermagnetism and surface reconstruction}
\label{SAM_and_reconstruction}
\change{To make progress on the question of how surface reconstruction might affect the surface induced altermagnetism, we adopt a simplified picture in which the reconstructed crystal motifs at the surface do not further lower the symmetries compared to the ideal surface termination. Under this assumption, an $N\times N$ surface reconstruction, in which the surface periodicity is enlarged, retains the same point symmetries of the underlying Bravais lattice as the unreconstructed surface. Together with the assumption that the surface motif remains unchanged in terms of point group symmetries, such reconstructions preserve the equivalence of the in-plane directions $x$ and $y$, and therefore also preserve the $90^\circ$ rotational symmetries that are crucial for $d$- and $g$-wave surface altermagnetism in the groups ${}^24/{}^1m{}^2m{}^1m$ and ${}^14/{}^1m{}^2m{}^2m$, respectively.
There also exist mismatched supercell reconstructions of the form $N\times M$, as famously exemplified by the Si(100) surface~\cite{Zangwill1988,Oura2003}, which exhibits a $2\times1$ surface reconstruction. Such reconstructions break the $90^\circ$ rotational symmetry by lifting the equivalence of the in-plane directions, while still preserving the $180^\circ$ rotational symmetry of the underlying Bravais lattice. In the case of a $g$-wave SAM, even within our simplified assumption that the symmetry of the surface motifs remain unchanged, the surface spin space group ${}^14{}^2m_{x/y}{}^2m_d$ is reduced to ${}^2m_x{}^2m_y{}^12$, yielding a $d$-wave SAM. We explicitly indicate the mirror generators here in order to make the symmetry-breaking pattern induced by the surface reconstruction transparent.
In the $d$-wave SAM cases of ${}^24{}^2m_{x/y}{}^1m_d$ and ${}^24{}^1m_{x/y}{}^2m_d$, the reconstruction yields either a $d$-wave surface altermagnet or an uncompensated surface, respectively, because only the $m_{x/y}$ mirror symmetries survive in the reconstructed system. In principle, certain motif reconstructions can also break the $[C_2||C_{2z}]$ symmetry characteristic of SD-SAM. This would enable a reconstruction-induced transition from the surface spin point group ${}^1m{}^2m{}^22_z$ to ${}^2m$. The latter corresponding to a $d$-wave SAM with \textit{nonrelativistic} spin-polarized surface bands compared to the spin-degenerate surface states of the unreconstructed SD-SAM.}

\newpage

\section{List of surface altermagnet candidates in the MAGNDATA materials database}

\label{fulllist}

\setlength{\LTcapwidth}{\textwidth}
\begin{longtable}{c c l c c l c}
\caption{Results of the MAGNDATA \cite{MAGNDATA, Gallego2016a} database screening for three-dimensional AFM with altermagnetic surfaces. Material entries with only AFM or uncompensated surfaces are not included. All uncompensated surfaces for the listed material entries are omitted. The first column entitled ``No.'' enumerates the entries. The second column called ``Material entry'' provides the chemical formula of the compounds. The third column ``Index'' gives the MAGNDATA entry index. The forth column called ``Normals'' shows the normal vectors describing all compensated surfaces in the material. {These normals are written in the coordinates spanned by the lattice vectors, with the crystal-setting convention used in MAGNDATA \cite{MAGNDATA, Gallego2016a}}. The fifth column ``Surface spin Laue group'' denotes the surface spin Laue group for SAM and SD-SAM surfaces and the crystallographic surface Laue group for AFM surfaces. The surface spin Laue groups are given in Litvin's notation \cite{Litvin1977}. SAM surface groups are colored blue, SD-SAM are colored black and AFM surface groups are colored lightgrey. {For the surface spin Laue groups corresponding to AFMs (gray font), we list here a crystallographic point group $\mathbf{G}$, in such a way that the spin Laue group is given by $\mathbb{Z}_2^{[C_{2\perp}||E]}\times \mathbf{G}=[E||\mathbf{G}]+[C_{2\perp}||\mathbf{G}]$}.  The sixth column entitled ``Surface character'' gives the partial-wave character for SAMs, and reads AFM or SD-SAM in the other cases. \textcolor{black}{In the final column entitled ``Terrace domains'' we indicate by ``yes/no'' whether or not there are different terminations for altermagnetic surfaces that with exactly opposite spin splitting.}}\label{tab:table}\\
\toprule
\multicolumn{1}{c}{No.} & \multicolumn{1}{c}{Material} & \multicolumn{1}{c}{Index} & \multicolumn{1}{c}{Normal} & \multicolumn{1}{c}{Surface spin Laue group} & \multicolumn{1}{c}{Partial wave} &\multicolumn{1}{c}{Terrace domains} \\
\midrule
\endfirsthead
\toprule
\multicolumn{1}{c}{No.} & \multicolumn{1}{c}{Material} & \multicolumn{1}{c}{Index} & \multicolumn{1}{c}{Normal} & \multicolumn{1}{c}{Surface spin Laue group} & \multicolumn{1}{c}{Partial wave} &\multicolumn{1}{c}{Terrace domains} \\
\midrule
\endhead
\addlinespace[1.2ex]
\multirow{2}{*}{1} & \multirow{2}{*}{$\mathrm{AgNiO_{2}}$} & \multirow{2}{*}{1.50} & $[001]$ & \textcolor{lightgrey}{$\bar{1}$} & AFM &\multirow{2}{*}{yes} \\
 &  &  & $[010]$ & ${}^22_{z}/ {}^2m_z$ & SD-SAM& \\
\addlinespace[1.2ex]
\specialrule{0.4pt}{2pt}{2pt}
\addlinespace[1.2ex]
\multirow{3}{*}{2} & \multirow{3}{*}{$\mathrm{AgRuO_{3}}$} & \multirow{3}{*}{0.733} & $[110]$ & ${}^22_{z}/ {}^2m_z$ & SD-SAM &\multirow{3}{*}{no} \\
 &  &  & $[100]$ & ${}^22_{z}/ {}^2m_z$ & SD-SAM& \\
 &  &  & $[010]$ & ${}^22_{z}/ {}^2m_z$ & SD-SAM& \\
\addlinespace[1.2ex]
\specialrule{0.4pt}{2pt}{2pt}
\addlinespace[1.2ex]
\multirow{5}{*}{3} & \multirow{5}{*}{$\mathrm{Ba_{2}FeSi_{2}O_{7}}$} & \multirow{5}{*}{1.641} & $[001]$ & \textcolor{blue}{${}^2m {}^2m {}^1m$} & $d$-wave &\multirow{5}{*}{yes} \\
 &  &  & $[010]$ & \textcolor{lightgrey}{$\bar{1}$} & AFM& \\
 &  &  & $[110]$ & \textcolor{lightgrey}{$2/m$} & AFM& \\
\bottomrule
\end{longtable}

\end{document}